\documentclass[11pt]{article}
\usepackage{epsfig,amscd,amssymb}
\begin{document}

\def\wgta#1#2#3#4{\hbox{\rlap{\lower.35cm\hbox{$#1$}}
\hskip.2cm\rlap{\raise.25cm\hbox{$#2$}}
\rlap{\vrule width1.3cm height.4pt}
\hskip.55cm\rlap{\lower.6cm\hbox{\vrule width.4pt height1.2cm}}
\hskip.15cm
\rlap{\raise.25cm\hbox{$#3$}}\hskip.25cm\lower.35cm\hbox{$#4$}\hskip.6cm}}
\def\wgtb#1#2#3#4{\hbox{\rlap{\raise.25cm\hbox{$#2$}}
\hskip.2cm\rlap{\lower.35cm\hbox{$#1$}}
\rlap{\vrule width1.3cm height.4pt}
\hskip.55cm\rlap{\lower.6cm\hbox{\vrule width.4pt height1.2cm}}
\hskip.15cm
\rlap{\lower.35cm\hbox{$#4$}}\hskip.25cm\raise.25cm\hbox{$#3$}\hskip.6cm}}

%
%
%
%

\title{Integrable quantum field theories with $OSP(m/2n)$ symmetries}

\author{Hubert Saleur and Birgit Wehefritz-Kaufmann\\
\smallskip\\
Department of Physics and Astronomy\\
University of Southern California\\
Los Angeles, CA 90089}

\maketitle

\begin{abstract}
\noindent We conjecture the factorized scattering description for 
$OSP(m/2n)/OSP(m-1/2n)$ supersphere sigma models and  $OSP(m/2n)$ Gross Neveu
models. The non unitarity of these field theories 
translates into a lack of `physical unitarity' of the 
S matrices, which are instead 
unitary  with respect to the non-positive scalar product inherited
from the orthosymplectic structure. Nevertheless, we find that 
formal thermodynamic Bethe ansatz
calculations appear meaningful, reproduce the correct central charges, and agree with
perturbative calculations. This paves the way to a more thorough study
of these and other models with supergroup symmetries using the S matrix approach.

\end{abstract}

\section{Introduction}

The field theory approach to phase transitions in disordered systems 
has realized major progress over the last few years, thanks to an ever deeper understanding of
two dimensional field theories. Conformal invariance, combined with  elegant reformulations using supersymmetry 
\cite{Efetov,LFSG,B}, and a greater control  of non unitarity issues
\cite{MCW,GLL,Z}, now severely constrains the possible fixed points
\cite{Victor,John}. In some simple cases, perturbed conformal field
theory, combined  with the use of current algebra symmetries, has even
led to complete solutions  \cite{GLL,Miraculous}. Some of the models of interest in the context of disordered systems have also appeared independently in string theory \cite{Be,W}, and more progress can only be expected from the cross fertilization between these two areas. 

Remarkably, the chief non perturbative method, the integrable
approach, has not been pushed very far to study these models. This is
a priori surprising. For instance, several disordered problems involve
variants of the $OSP(m/2n)$ Gross Neveu model, which formally appears
just as integrable as its well known $O(N)$ counterpart. The standard
way of proceeding to study such a model would be to determine  its S matrix, and then use the thermodynamic Bethe ansatz and form-factors to calculate physical properties. This approach was pioneered in the elegant papers \cite{MS,CHMP}, and revived in \cite{BL}, but so far the subject was only touched upon in our opinion; for instance,  although the S matrix of the $OSP(2/2)$ Gross Neveu model has been conjectured \cite{BL}, no calculation to justify this conjecture has been possible. Super sigma models have also been tackled, this time in the context of string theory \cite{Sethi}, but there 
again results have only been very partial, and the S matrix approach even less developed than for super Gross Neveu models.

The main reasons for this unsatisfactory situation seem
technical. While there has been  tremendous  progress in the
understanding of the sine-Gordon model and the $O(3)$ sigma models -
the archetypes of  integrable field theories -  models based on other Lie algebras are only partially understood (see \cite{SWK,Paul} for some recent progress), and the situation becomes even more confusing when it comes to super-algebras.  
One of the main difficulties in understanding these theories is physical, and related with a general  lack of unitarity - a feature that is natural from the disordered condensed matter point of view, but confusing at best from a field theory stand point. Another difficulty is 
simply the complexity of the Bethe ansatz for higher rank algebras, in
particular super algebras. While these equations can be written
sometimes (see the recent  
recent tour de force \cite{MartinsRamos}), finding the pattern of solutions - the generalized string hypothesis - is a daunting task even for the 
trained expert \cite{Essler}.  

Integrable field theories and lattice models go hand in hand, and the foregoing confusion 
seems to extend to spin chains based on superalgebras. Although the formalism is by now well in place to write the integrable Hamiltonians, 
their continuum limit is not well understood. In the case of ordinary algebras for instance, it is known that this continuum limit 
is a Wess Zumino model on the group: whether this is true or not for
superalgebras has been  a matter of some debate  \cite{MarNien}. 
Note that in some cases, the super spin chain is better understood 
than the field theory: this is the case for instance of the 
$sl(2/1)$ spin chain of \cite{GLR, Read}
in the spin quantum Hall effect, whose relation to the traditional (super) Yang Baxter formalism is also not understood at the present time. 

Our purpose in this paper is to develop  the integrable approach for the case of $OSP(m/2n)$ field theories. We will discuss two kinds of models, the supersphere sigma-models, and the Gross Neveu models, mostly for algebras $OSP(1/2n)$. In each case, we will conjecture a scattering theory, whose striking feature will be the lack of unitarity of the $S$ matrices, as a result of  the super group symmetry.
We will argue that formal thermodynamical calculations do make sense nevertheless, and illustrate this point for both types of models.

\section{Algebraic Generalities}

There are two basic integrable models with $O(N)$ symmetry, the Gross Neveu model and the sphere sigma model $S^{N-1}=O(N)/O(N-1)$. Once their integrability is proven, the scattering theory is determined by implementing the action of the symmetry on the space of particles, 
and by  requiring  factorization. This is not always an obvious task, because of issues of bound states and charge fractionalization. For instance, the scattering theory for the $O(2P+1)$ Gross Neveu model
was completed only very recently \cite{FS}. However, the scattering of particles in the defining representation has been known for a long time \cite{ZZ}, and this is where we would like to start here. 

Scattering matrices with $O(N)$ symmetry can generally be written in terms of three independent tensors
\begin{equation}
\check{S}_{i_1j_1}^{j_2i_2}=\sigma_1 E+\sigma_2 P+\sigma_3 I \label{maini} 
\end{equation}
where  we have set
\begin{eqnarray}
E_{i_1j_1}^{j_2i_2}&=\delta_{i_1j_1}\delta^{i_2j_2}\nonumber\\
P_{i_1j_1}^{j_2i_2}&=\delta_{i_1}^{i_2}\delta_{j_1}^{j_2}\nonumber\\
I_{i_1j_1}^{j_2i_2}&=\delta_{i_1}^{j_2}\delta_{i_2}^{j_1}
\end{eqnarray}
corresponding to the  graphical representation in figure 1. 

\begin{figure}
\centerline{\epsfxsize=4.0in\epsffile{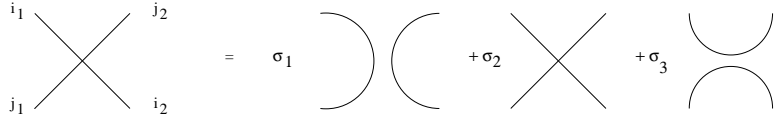}}
\bigskip\bigskip
\caption{Graphical representation of the invariant tensors appearing
  in the $S$ matrix. }
\end{figure}

We are interested here in  models for which none of the amplitudes
vanish. Specifically, for $N$ a positive integer, there are generically two known 
models whose scattering  matrix for the vector representation has the form  (\ref{maini}), with none of the $\sigma_i$'s vanishing. They are given by 
\begin{eqnarray}
\sigma_1&=&-{2i\pi\over (N-2)(i\pi-\theta)}~\sigma_2\nonumber\\
\sigma_3&=&-{2i\pi\over (N-2)\theta}~\sigma_2\label{mainii}
\end{eqnarray}
with two possible choices for $\sigma_2$:
\begin{equation}
\sigma_2^{\pm}(\theta)={\Gamma\left(1-{\theta\over 2i\pi}\right)\over \Gamma\left({\theta\over 2i\pi}\right)}
{\Gamma\left({1\over 2}+{\theta\over 2i\pi}\right)\over \Gamma\left({1\over 2}-{\theta\over 2i\pi}\right)}
{\Gamma\left(\pm{1\over N-2}+{\theta\over 2i\pi}\right)\over  \Gamma\left(1\pm{1\over N-2}-{\theta\over 2i\pi}\right)}
{\Gamma\left({1\over 2}\pm{1\over N-2}-{\theta\over 2i\pi}\right)\over \Gamma\left({1\over 2}\pm{1\over N-2}+{\theta\over 2i\pi}\right)}\label{mainiii}
\end{equation}
The factor $\sigma_2^+$ does not have  poles in the physical strip for $N\geq 0$, and the corresponding S matrix for $N\geq 3$  is believed to describe the $O(N)/O(N-1)$ 
sphere ($S^{N-1}$) sigma model. The factor $\sigma_2^-$ does not have poles in the physical strip for $N\leq 4$. For $N>4$, it describes the 
scattering of vector particles in $O(N)$ Gross Neveu model. Recall that for $N=3,4$ the vector particles are unstable and disappear from the spectrum, that contains only kinks. Some of these features are illustrated for convenience in figure 2.

\begin{figure}
\centerline{\epsfxsize=4.0in\epsffile{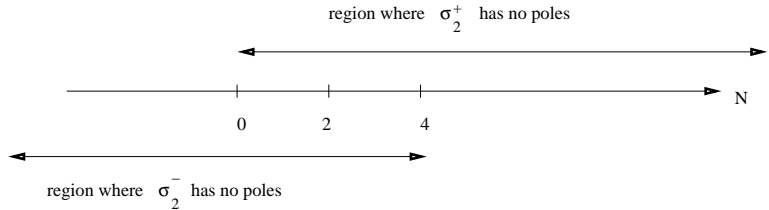}}
\bigskip\bigskip
\caption{Pole structure of $\sigma_2$ as a function of $N$. }
\end{figure}

Note that at vanishing rapidity, the scattering matrix reduces to $\check{S}(\theta=0)=\mp I$. This is in agreement with the fundamental particles being bosons in the sigma model , and fermions in the Gross Neveu model \cite{KT}.

Our next step is to  try to define models for which $N<1$, in particular $N=0$, or $N$ negative. 
A similar question has been tackled by Zamolodchikov \cite{Zamopoly} under the condition that 
 particles be ``impenetrable'', 
that is  $\sigma_1=0$. The (standard) procedure he used was to study the algebraic relations 
satisfied by the objects  $E,I$  for integer $N$, extend these relations to arbitrary $N$,
and find objects (not necessarily $N\times N$ matrices) satisfying them. In technical terms, 
the algebraic relations turned out to be the defining ones for the Temperley Lieb algebra \cite{Martin}, for which 
plenty of representations were known. 
 The most interesting $N=0$ case (corresponding to polymers) could then be studied 
 using the 6-vertex model representation. It could  also be studied  using algebras $OSP(2n/2n)$, or algebras $GL(n/n)$.

In trying to address the same question for models where $\sigma_1\neq 0$, it is natural to  set up the problem in algebraic terms again. The objects $E,P,I$ can be understood as providing a particular representation of  the following Birman Wenzl \cite{BW} algebra, defined by
 generators $E_i,P_i$, $i=1,\ldots$ and  relations
\begin{eqnarray}
P_iP_{i\pm 1}P_i&=&P_{i\pm 1}P_iP_{i\pm 1}\nonumber\\
P_i^2&=&1\nonumber\\
\left[P_i,P_j\right]&=&0,~~~|i-j|\geq 2,\label{BWi}
\end{eqnarray}
together with 
\begin{eqnarray}
E_iE_{i\pm 1}E_i&=&E_i\nonumber\\
E_i^2&=&NE_i\nonumber\\
\left[E_i,E_j\right]&=&0,~~~|i-j|\geq 2,\label{TL}
\end{eqnarray}
and
\begin{eqnarray}
P_iE_i&=E_iP_i=&E_i\nonumber\\
E_iP_{i\pm 1}P_i&=P_{i\pm 1}P_iE_{i\pm 1}=&E_iE_{i\pm 1}\label{BWii}
\end{eqnarray}
 These relations can be interpreted graphically as in figure 3; operators $E$ define a sub Temperley Lieb algebra \cite{Martin}.

\begin{figure}
\centerline{\epsfxsize=3.5in\epsffile{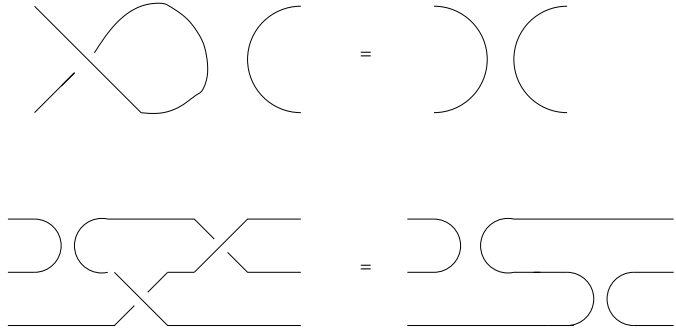}}
\bigskip\bigskip
\caption{Graphical representation for the defining relations of the BW
  algebra.}
\end{figure}

The natural extension of what was done say for polymers would be to look for vertex representations of the Birman Wenzl algebra. However, this does not seem possible. The point is that the full Birman Wenzl algebra has two parameters, and the representation furnished say by the spin one vertex model
will have, for instance, that $P_i\neq P_i^{-1}$. This is a  property natural from the knot theory framework where this algebra comes from, but disastrous for the construction of physical $S$ matrices, where  particles cannot  ``go under'' another.  Extending the definition of the S matrix to arbitrary values of $N$ thus seems problematic. 

It is  easy nevertheless to extend it to negative integer values of $N$. Indeed, the Birman Wenzl algebras 
arise from representation theory of $O(N)$, and most of the properties of these algebras generalize to the superalgebras $OSP(m/2n)$\cite{Frappat}.  Instead of the vector representation of $O(N)$, take the vector  representation of the orthosymplectic algebra, of dimensions $(m,2n)$. For $m\neq 2n$, the tensor product with itself gives rise to three representations. Taking $I$  as the identity, $E$ as $(m-2n)$ times the projector on the identity representation, and $P$ as the graded permutation operator (the extension to the case  $m=2n$ is easy), it can be checked indeed that the relations (\ref{BWi}),(\ref{TL}),(\ref{BWii}) are obeyed 
 with $N=m-2n$.  More explicitly, in the usual case, the matrix elements of $E$ are obtained by
contracting the ingoing and outgoing indices using the unit matrix. In
the $OSP$ case, they are obtained similarly by contracting indices
using the defining form of the $OSP$ algebras
\begin{equation}
J=\left(\begin{array}{ccc}
I_m&0&0\\
0&0&-I_n\\
0&I_n&0\end{array}\right)
\end{equation}
In formulas, we set $\bar{i}=i,i=1,\ldots m$, $\bar{i}=n+i,
i=m+1,\ldots,m+n$, $\bar{\bar{i}}=i$. We set  $x(i)=1, i=m+1,\ldots,m+n$, $x(i)=0$
otherwise, so $p(i)=x(i)+x(\bar{i})$. One has then
\begin{equation}
E_{i_1j_1}^{j_2i_2}=\delta_{i_1,\bar{j}_1}\delta^{i_2,\bar{j}_2}
(-1)^{x(i_1)}(-1)^{x(i_2)}
\end{equation}
while the graded permutation operator is of course given by
\begin{equation}
P_{i_1j_1}^{j_2i_2}=(-1)^{p(i_1)p(j_1)}\delta_{i_1}^{i_2}\delta_{j_1}^{j_2}
\end{equation}
This realization of Birman Wenzl algebras was first mentioned  in the very interesting paper \cite{MarNien}. It thus follows that the natural ortho-symplectic generalization of the $\check{S}$ matrix of the $O(N)$ Gross Neveu  model (or sphere sigma model) 
does provide a solution of the Yang Baxter equation, and realizes algebraically  the continuation to values of $N$ equal to zero or 
negative integers. Let us now discuss how meaningful this can be physically.

For this, let us recall some basic features about Yang-Baxter versus graded Yang-Baxter. 
In all cases, the Yang Baxter formalism deals with 
two related objects that are usually  called $R,\check{R}$ in a general context,  $S,\check{S}$ in the context of scattering theory, and differ by some (graded) permutations. 

In the ordinary case,
we reserve the unchecked symbol  to the matrix obeying
$R_{12}(u-v)
R_{13}(u)R_{23}(v)=R_{23}(v)R_{13}(u)R_{12}(u-v)$, where
$u,v$ are spectral parameters. The equivalent of this relation for the superalgebra case is the graded Yang Baxter equation, and it
involves signs \cite{Review}:
\begin{eqnarray}
R_{i_1 i_2}^{k_1k_2}(u-v)R_{k_1i_3}^{j_1k_3}(u)R_{k_2k_3}^{j_2j_3}(v)
(-1)^{p(i_1)p(i_2)+p(k_1)p(k_3)+p(k_2)p(k_3)}=\nonumber\\
R_{i_2i_3}^{k_2k_3}(v)R_{i_1k_3}^{k_1j_3}(u)R_{k_1k_2}^{j_1j_2}(u-v)
(-1)^{p(i_2)p(i_3)+p(i_1)p(k_3)+p(k_1)p(k_2)}
\end{eqnarray}
where $p(k)=1,0$ is the parity of the $k$ coordinate. These signs occur   because, in the graded tensor product formalism, $R_{13}$ acts on the first and third components, hence giving rise to potential minus signs when
commuting through the elements of the second component. An ordinary (super) R matrix does not solve the graded (ordinary)  Yang Baxter equation. However, if $R$ does solve the graded Yang-Baxter equation, the object $\tilde{R}_{ij}^{kl}\equiv (-1)^{p(i)p(j)} R_{ij}^{kl}$ solves then the ordinary Yang-Baxter equation, so it is easy to go from one point of view to the other. 

In the ordinary case, one can also consider the object $\check{R}=PR$, $P$ the permutation operator: this is what we gave in formula (\ref{maini}) for the case $N$ a positive integer. It satisfies 
a different relation, $\check{R}_{12}(u-v)\check{R}_{23}(u)\check{R}_{12}(v)=\check{R}_{23}(v)\check{R}_{12}(u)\check{R}_{23}(u-v)$.  Observe 
that this relation now involves only neighboring spaces in the tensor product, and thus is insensitive to grading. If $R$ were to solve the graded Yang Baxter equations instead, 
the same relation would be obeyed  by the matrix $\check{R}=PR$, where now $P$ is the graded permutation operator. Whether $R$ satisfies the ordinary or the graded Yang-Baxter equation, it follows that matrices $\check{R}$ do satisfy the same equation. Conversely, a solution of 
$\check{R}_{12}\check{R}_{23}\check{R}_{12}=\check{R}_{23}\check{R}_{12}\check{R}_{23}$ 
 can  be interpreted as arising from a graded or a non graded structure. 
 The graded Yang Baxter equation appears more as an esthetically appealing object than a fundamental one. It is especially nice because it admits a classical limit, and fits in the general formalism of quantum super groups \cite{Reviewi}. 

In the context of  scattering theories, which are our main interest here, it is convenient to define the $S$ matrix  through the 
 Fadeev Zamolodchikov algebra \cite{Smirnovbook}. Theories based on
 supergroups will have a spectrum of particles containing both bosons
 and fermions. Their creation and annihilation  operators  will be
 denoted $Z^{(\dagger)}$, and obey for instance $Z^\dagger_i(\theta_1)Z^\dagger_j(\theta_2)=(-1)^{p(i)p(j)} S_{ij}^{kl}(\theta_1-\theta_2) Z^\dagger_l(\theta_2)Z^\dagger_l(\theta_1)$.
The consistency of these relations requires that  $S$ satisfies the
graded Yang Baxter equation, or, equivalently, that $\tilde{S}$
satisfies the ordinary Yang Baxter equation.
Amplitudes of physical processes are then derived  in
the usual way. An important feature is that the monodromy matrix,
which describes scattering of a particle through others, is built out
of $\tilde{S}$ like in the non graded case (the same thing happens for
integrable lattice models \cite{Reviewi}). 

Taking therefore our $OSP$ $\check{S}$ matrix, and the $S$ matrix that follows from it, $S=\sigma_1E+\sigma_2I+\sigma_3P$, it is natural to ask about the physical meaning of these amplitudes. This reveals some surprises.  
Crossing and unitarity are well implemented in the cases when the particles are bosons or fermions. Mixing the two kinds does not seem, a priori,  to give rise to any difficulty. For instance the relation $S(\theta)S(-\theta)=\check{S}(\theta)\check{S}(-\theta)=I$ holds  in the graded case  with proper choice of normalization factors. It will turn out however that in the graded case, 
the $S$ matrix is, as a matrix, {\bf not unitary} \footnote{This is a stronger
violation of unitarity than in cases like the Lee-Yang singularity,
where $SS^\dagger=1$ still holds, but unphysical signs appear in $S$
matrix residues. For a thorough discussion of unitarity issues see \cite{rsosnonsense}.}. It is thus difficult  to interpret our $S$ matrices in terms of 
a `physical' scattering. The most useful way to think of  the $S$
matrices 
 will  probably be as an object describing the monodromy of wave functions, like in imaginary Toda theories \cite{SWK,TakacsWatts}. Crossing follows then from $\check{S}(i\pi-\theta)=\sigma_1 (\theta)I+\sigma_2 (\theta)P+\sigma_3(\theta) E$,
with an obvious graphical interpretation, and charge conjugation being defined through the defining form of the $OSP$ algebra.

Leaving aside the unitary difficulty, the usual formal procedure thus  selects once again the factors $\sigma_2^\pm$ as minimal prefactors, with the continued values $N=m-2n$. The question is then to  establish the relation what  field theory, if any.

Obvious candidates are  the $OSP(m/2n)$ Gross Neveu model with action
(in all this paper, normal ordering is left implicit)
\begin{equation}
S=\int {d^2x\over 2\pi}  \left[\sum_{i=1}^{m}\psi^i_L\partial\psi^i_L+\psi^i_R\bar{\partial}\psi^i_R+\sum_{j=1}^{n}
\beta_L^j\partial\gamma_L^j+\beta_R^j\bar{\partial}\gamma_R^j
+g
\left(\psi_L^i\psi_R^i+\beta_L^j\gamma_R^j-\gamma_L^j\beta^j_R\right)^2\right]
\end{equation}
where the $\psi$ are Majorana fermions of conformal weight $1/2$, and
the $\beta\gamma$ are bosonic ghosts of weight $1/2$ as
well. Perturbative calculations of the beta function \cite{B,Wegner}
suggest that this model behaves like the continuation of the $O(N)$
Gross Neveu model to the value $N=m-2n$. Similarly, the natural
generalization of the sphere sigma model is a super sphere sigma
model, which can be described as the coset $OSP(m/2n)/OSP(m-1/2n)$
. There again, perturbative beta functions do match. It is therefore
natural to expect that the S matrices built on $OSP(m/2n)$ will
describe, depending on the prefactor $\sigma_2^\pm$, these two models
in the appropriate physical regimes. This will be discussed in the
next section.

\section{The $OSP(1/2)$ sigma model S matrix}

\subsection{The S matrix}

To make things more concrete,  let us discuss the case $N=-1$, and its realization using $OSP(1/2)$. Instead of the Gross Neveu model, it will turn out to be easier to study the equivalent of the sigma model, because of its relation with the  $a_2^{(2)}$ Toda theory and spin chain.

The solution of the graded Yang Baxter equation relevant here is the well known $OSP(1/2)$ one, given by
\begin{equation}
R_{OSP(1/2)}= {1\over 1-3{\theta\over 2i\pi}} \left[P+{3\theta\over 2i\pi}I+{\theta\over i\pi-\theta}E\right]
\end{equation}
where we have chosen the  normalization factor for later purposes, $I$ is the identity. Denote the basis vectors in the fundamental representation of $OSP(1/2)$ as $b,f_1,f_2$. The operator $E$ is given  by the matrix
\begin{equation}
E=\left(\begin{array}{ccc}
1 & -1 & 1\\
1 & -1& 1\\
-1 & 1 &-1\end{array}\right)
\end{equation}
in the subspace spanned by $(b,b),(f_1,f_2),(f_2,f_1)$ in that order, $E=0$ otherwise. In that same subspace, the graded permutation operator 
reads
\begin{equation}
P=\left(\begin{array}{ccc}
1 & 0 & 0\\
0 & 0& -1\\
0 & -1 & 0\end{array}\right)
\end{equation}
The operators $E,P$ satisfy the defining relations of the Birman Wenzl algebra with $N=-1$.  

The non graded $\check{R}$ matrix meanwhile reads
\begin{equation}
\check{R}_{OSP(1/2)}={1\over 1-3{\theta\over 2i\pi}} \left[I+{3\theta\over 2i\pi}P+{\theta\over i\pi-\theta}E\right]\label{rosp}
\end{equation}

Let us now discuss the issue of unitarity. While
$R(\theta)R(-\theta)=\check{R}(\theta)\check{R}(-\theta)=1$, $R$,
$\check{R}$, and $\tilde{R}$ as matrices,  are unitary only 
with respect to an indefinite metric induced by the supergroup structure. Explicitly, one has for instance
\begin{equation}
\check{R}_{bb}^{bb}\check{R}_{bb}^{*bb}-\check{R}_{bb}^{f_1f_2}\check{R}_{bb}^{*f_1f_2}-
\check{R}_{bb}^{f_2f_1}\check{R}_{bb}^{*f_2f_1}=1
\end{equation}
and in fact $\check{R}$ conserves a scalar product that 
allows for negative norm square states $<ff|ff>=-1$, all others
equal to $+1$. It is  well known indeed  \cite{Scheunert} that  the structure
  of $OSP(1/2)$ is not compatible with a positive scalar product. The
  mere presence of supergroup symmetry leads necessarily to the
  existence of negative norm-square states, and therefore to unitarity
  problems.

%
%
The resulting scattering matrix is therefore non unitary, in the usual
sense.  This is a consequence of the orthosymplectic  
supergroup symmetry, and  originates physically in the non unitarity
of the field theory described by the $S$ matrix. This does not prevent one from using the $S$ matrix at least
to describe the monodromy of the wave functions, as we will do in the
section devoted to TBA. Similarly, this $S$ matrix could also be used
to describe aspects of the finite size spectrum \cite{rsosnonsense,TakacsWatts}.

An intriguing remark is that, although the matrix $\tilde{R}$ is not unitary, its
eigenvalues happen to be complex numbers of modulus one (the same hold
for $R$ and $\check{R}$)\footnote{We
 thank G. Takacs for suggesting this may be the case.}, and there are reasons
to believe that this is true for the eigenvalues of the monodromy
matrices involving an
arbitrary number of particles. This means that the non unitarity
situation is not as stringent as say in the $a_2^{(1)}$ case
\cite{SWK}, and that, for instance, the spectrum of the theory in
finite size will be real.  

Let us now consider the `scattering' theory that is the continuation of the sphere sigma model to $N=-1$: we take the $OSP(1/2)$ realization, 
and as a prefactor  $\sigma_2^+$. It then  turns out 
that the $S$  matrix is identical to the one of the $a_2^{(2)}$ Toda theory for a particular value of the coupling constant! This  
will allow us to explicitly perform the TBA, and identify the scattering theory indeed. While we were carrying out these calculations, we 
found out two papers where the idea has been carried out to some extent already: one by Martins \cite{Martins}, and one by Sakai and Tsuboi \cite{ST}. Our 
approach has little overlap with these papers, and stems from our earlier work on the $a_2^{(1)}$ theory instead. 

To proceed, we now discuss the $a_2^{(2)}$ Toda theory in more details. 

\subsection{A detour through $a_2^{(2)}$}

This theory has action
\begin{equation}
S={1\over 8\pi} \int dx dy
\left[(\partial_x\Phi)^2+(\partial_y\Phi)^2+\Lambda(2e^{-{i\over
      \sqrt{2}}
\beta\Phi}+
e^{i\sqrt{2}\beta\Phi})\right] \label{Toda}
\end{equation}
The conformal weight of the first field is $\Delta_1={\beta^2\over 4}$, while the one
of the second is $\Delta_2=\beta^2$. The dimension of $\Lambda$ is
such that $[\Lambda]^3 L^{-2h_2-4h_1}=L^{-6}$, so $[\Lambda]=L^{-2}
L^{2h_2+4h_1\over 3}$, i.e. $[\Lambda]=L^{\beta^2-2}$, and the
``effective'' dimension  (i.e. twice the conformal weight) of the perturbation is $d=\beta^2$.  

The domain we shall be interested in primarily corresponds to
$\beta^2\geq 1$. 
We will  parameterize  
\begin{equation}
\beta^2=2{t-1\over t}\label{dimparam}
\end{equation}
so $h_1={\beta^2\over 4}={t-1\over 2t}$, $[\Lambda]=L^{-2/t}$. The case $t=2$ corresponds to $h_1={1\over
  4}$,
and the limit $t\rightarrow \infty$ to $h_1={1\over 2}$. 

The massless or massive nature of the theory depends on the sign of
$\Lambda$ and on the value of  $\beta^2$  \cite{Fateev}. For $\beta^2\leq 1$, the
theory is massive for $\Lambda<0$, but for the region we are
interested in, $\Lambda>0$ is required, and we will restrict to this
in the following.

In the $t\in [2,\infty]$ domain, the scattering matrix has been first  conjectured by Smirnov \cite{Smirnov}. The spectrum does not contain any bound states,
and is simply made of solitons with topological charges $\pm 1,0$
(where the topological charge is defined as $q={1\over 2\sqrt{2}\pi\beta}\int
\partial_x\phi$).
 The relation between the mass of the solitons and
the coupling constant reads \cite{Fateev}
\begin{equation}
\Lambda^3=- {1\over 16\pi^3} {\Gamma^2(\beta^2/4)\Gamma(\beta^2)\over
  \Gamma^2(1-\beta^2/4)\Gamma(1-\beta^2)}
\left[
{\pi M\over \sqrt{3}\Gamma(1/3)}
 {\Gamma\left({2\over
  3(2-\beta^2)}\right)\over \Gamma\left({\beta^2\over
  3(2-\beta^2)}\right)}\right]^{3(2-\beta^2)}
\end{equation}
Near $\beta^2=2$, which will turn out to be  the point with $OSP(1/2)$  symmetry, setting $\beta^2=2-\epsilon$, one has
$\Lambda^3\propto \epsilon M^{3\epsilon}$.

 The $\check{S}$ matrix is proportional to the 
$\check{R}$ matrix of the Izergin Korepin model \cite{Kolya}. Although this may seem
laborious, 
we will write it explicitly here. Introducing the parameter 
\begin{equation}
\xi={2\over 3}{\pi \beta^2\over 2-\beta^2}
\end{equation}
and the variables $\lambda=e^{-2\pi\theta/5\xi}$,
$p=e^{i\pi/2}e^{i\pi/3\xi}$,
we write 
$$
\check{S}=\Sigma_0 {1\over \lambda^5p^5-\lambda^{-1}p^{-5}+p^{-1}-p}~\check{R}
$$
with \cite{Kolya, Warnaar}
\begin{eqnarray}
\check{R}_{11}^{11}&=&\check{R}_{-1,-1}^{-1,-1}=\lambda^5p^5-\lambda^{-1}p^{-5}+p^{-1}-p\nonumber\\
\check{R}_{1,0}^{0,1}&=&\check{R}_{0,1}^{1,0}=\check{R}_{0,-1}^{-1,0}=\check{R}_{-1,0}^{0,-1}=\lambda p^3-\lambda^{-1}p^{-3}+p^{-3}-p^3\nonumber\\
\check{R}_{1,-1}^{-1,1}&=&\check{R}_{-1,1}^{1,-1}=\lambda p-\lambda^{-1}p^{-1}+p^{-1}-p\nonumber\\
\check{R}_{00}^{00}&=&\lambda p^3-\lambda^{-1}p^{-3}+p^{-3}-p^3+p^{-1}-p+p^5-p^{-5}\nonumber\\
\check{R}_{01}^{01}&=&\check{R}_{-1,0}^{-1,0}=\lambda(p^5-p)+p^{-1}-p^{-5}\nonumber\\
\check{R}_{0,0}^{-1,1}&=&\check{R}_{-1,1}^{0,0}=\lambda(p^4-1)+1-p^4\nonumber\\
\check{R}_{1,0}^{1,0}&=&\check{R}_{0,-1}^{0,-1}=\lambda^{-1}(p^{-1}-p^{-5})+p^5-p\nonumber\\
\check{R}_{1,-1}^{0,0}&=&\check{R}_{0,0}^{1,-1}=\lambda^{-1}(1-p^{-4})+p^{-4}-1\nonumber\\
\check{R}_{1,-1}^{-1,1}&=&\lambda(p^5-p-p^3+p^{-1})+p^3-p^{-5}\nonumber\\
\check{R}_{1,-1}^{1,-1}&=&\lambda^{-1}(p^{-1}-p^{-5}-p+p^{-3})+p^5-p^{-3}\label{schmat}
\end{eqnarray}
The normalization factor admits the representation
\begin{equation}
\Sigma_0=-\exp\left[i\int_{-\infty}^\infty  {d\omega\over \omega }e^{-3i\omega \theta/\pi}
{\sinh (3\omega) \cosh(3\omega(2\xi-\pi)/4)\over 
\sinh (3\omega\xi/2\pi)\cosh (3\omega/2)}
\right]
\end{equation}
It is equal to the amplitude for the scattering processes $11\rightarrow 11$. 

In the case  
$\xi\rightarrow \infty$, one checks that 
\begin{equation}
{1\over \lambda^5p^5-\lambda^{-1}p^{-5}+p^{-1}-p}\check{R}\longrightarrow  ~~~\check{R}_{osp(1/2)}
\end{equation}
(with $b\leftrightarrow 0$, $f_{1,2}\leftrightarrow \pm 1$) up to an irrelevant gauge transformation. Moreover, it turns out that 
\begin{equation}
\Sigma_0 ~{{3\theta\over 2i\pi}\over 1-{3\theta\over 2i\pi}}\longrightarrow \sigma_2^+
\end{equation}
or $\Sigma_0=\sigma_3^+-\sigma_2^+$ for $N=-1$, confirming the identification of the $a_2^{(2)}$ $\check{S}$ matrix in the limit $t\rightarrow\infty$ with the $OSP(1/2)$ 
``sphere sigma-model'' $\check{S}$ matrix.

This coincidence has a simple algebraic origin. Indeed  recall \cite{Efthimiou,Takacs}, that the $a_2^{(2)}$  Toda theory has symmetry $U_q(a_2^{(2)})$, $q=e^{i\pi/\beta^2}$. The 
Dynkin diagram for the algebra $a_2^{(2)}$   turns out to be almost identical to the one for the algebra $osp(1|2)^{(1)}$ \cite{Frappat}, as represented in figure 4, although in the latter case, one of the roots is fermionic, and therefore the basic relations
involve an anticommutator instead of a commutator.

\begin{figure}
\centerline{\epsfxsize=4.0in\epsffile{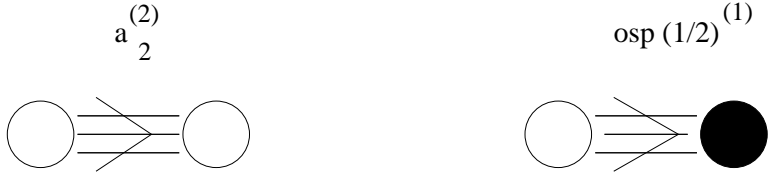}}
\bigskip\bigskip
\caption{Dynkin diagrams for $a_2^{(2)}$ and $osp(1/2)^{(1)}$. }
\end{figure}

It can be hoped that for some particular value of $q$, the $q$-deformation
of one algebra gives rise to the other, and this is what we shall 
now demonstrate - namely that there is a mapping 
between $U_q(a_2^{(2)})$ and $U(osp(1|2)^{(1)})$, for $q=i$. 
This should not come as a surprise, and has algebraic roots going back 
as far as \cite{Rittenberg}. For recent related works, see \cite{HS, BLsuper}.

Traditionally, the Cartan matrix of $a_2^{(2)}$ is written 
as $\left(\begin{array}{cc}
8 & -4\\
-4 & 2
\end{array}\right)$, and the commutation relations are
\begin{eqnarray}
\left[H_i,H_j\right]&=&0\nonumber\\
\left[H_i,E_j\right]&=&a_{ij}E_j\nonumber\\
\left[H_i,F_j\right]&=&-a_{ij}F_j\nonumber\\
\left[E_i,F_j\right]&=&\delta_{ij}{q^{H_i}-q^{-H_i}\over q_i-q_i^{-1}},~~~q_i=q^{a_{ii}/2}\label{genecom}
\end{eqnarray}
This means in particular that the generators $E_0,F_0,H_0$ satisfy a $U_{q^4}(a_1)$ algebra
\begin{eqnarray}
\left[H_0,E_0\right]&=&8E_0\nonumber\\
\left[H_0,F_0\right]&=&-8F_0\nonumber\\
\left[E_0,F_0\right]&=&{q^{H_0}-q^{-H_0}\over q^4-q^{-4}}
\end{eqnarray}
while the generators $E_1,F_1,H_1$ satisfy a $U_q(a_1)$ algebra
\begin{eqnarray}
\left[H_1,E_1\right]&=&2E_1\nonumber\\
\left[H_1,F_1\right]&=&-2F_1\nonumber\\
\left[E_1,F_1\right]&=&{q^{H_1}-q^{-H_1}\over q-q^{-1}}
\end{eqnarray}
The Cartan matrix of $osp(1|2)^{(1)}$ on the other hand reads usually $\left(\begin{array}{cc}
4 & -2\\
-2 & 1\end{array}\right)$. Commutation relations are similar to (\ref{genecom}), but involve anticommutators instead of commutators 
for the fermionic generators. The generators $e_0,f_0,h_0$ satisfy thus  a $a_1$ algebra
\begin{eqnarray}
\left[h_0,e_0\right]&=&4e_0\nonumber\\
\left[h_0,f_0\right]&=&-4f_0\nonumber\\
\left[e_0,f_0\right]&=&h_0
\end{eqnarray}
while for the generators $\psi_1^\dagger,\psi_1, h_1$ one has
\begin{eqnarray}
\left[h_1,\psi_1^\dagger\right]&=&\psi_1^\dagger\nonumber\\
\left[h_1,\psi_1\right]&=&-\psi_1\nonumber\\
\{\psi_1^\dagger,\psi_1\}&=&h_1
\end{eqnarray}
Taking $q=i$ for $U_q(a_2^{(2)})$ makes the subalgebra generated by $E_0,F_0,H_0$ a $U(a_1)$ algebra. The value $q=i$ for the 
other deformed $a_1$ was already observed in \cite{HS} to allow a simple relation with a fermionic algebra, a fact also used in mapping 
$U_i(a_1)$ onto a supersymmetric  ${\cal N}=1$ algebra. Here, observe that by setting $\psi_1^\dagger=q^{-(H_1+1)/2}E_1$ and $\psi_1=
q^{(H_1-1)/2}F_1/{(q+q^{-1})}$,
, one finds, {\sl for representations where $H_1$ is even} (the only ones of interest in our case), that, when $q\rightarrow i$, 
$$
\{\psi_1^\dagger,\psi_1\}= {H_1\over 2}
$$
in agreement with the anticommutation relation for $U(osp(1|2))$ if $h_1={H_1\over 2}$. The rest of the relations then are in complete 
agreement, up to some straightforward changes of normalization.

We conclude that, restricting to representations with $H_1$ even, the two algebras are isomorphic. Since this constraint is satisfied in the case at hand, the  $osp(1/2)^{(1)}$ symmetry of the $a_2^{(2)}$ Toda theory is thus explained.

\subsection{Thermodynamic Bethe ansatz}

Throughout this paper, we will use the thermodynamic Bethe ansatz
to calculate physical properties of our theory. It is a priori unclear
whether the method - which involves maximizing a free energy - makes
much sense in a theory whose Hamiltonian is not hermitian, but the
results we obtain seem perfectly meaningful, like in other similar
examples. Two additional remarks about the TBA are relevant. First,
the scattering matrix appearing in the auxiliary monodromy problem
(diagonalizing the matrix describing the effect of passing a particle
through the others) is not $S$ but $\tilde{S}$. This means that,
although the S matrices of the $osp^{(1)}$  and $a_2^{(2)}$ differ
because of the grading, the objects used in the TBA (like the
$\tilde{S}$ matrices) are identical, and known results about
$a_2^{(2)}$ Toda theories can be used. 
Second, one may worry that mixing bosons and fermions  could give rise to problems in applying the TBA. This is not quite so however. Most TBA's known so far - and the ones we will introduce here will be no exceptions - allow at most one particle in a state of a given rapidity. As discussed in Zamolodchikov \cite{oldtba}, this corresponds, in the diagonal case,  to having $S_{ii}^{ii}(0)=-(-1)^F$, where $F$ is the fermion number of particle $i$. 
In our case, we have $S_{ii}^{ii}(0)=\mp P_{ii}^{ii}$. For the supersphere sigma model, the particles with bosonic internal labels $i=1,\ldots, m$ will be bosons, so $P_{ii}^{ii}=(-1)^F$. For the super Gross-Neveu model, the particles with bosonic 
internal labels are now fermions, so $P_{ii}^{ii}=-(-1)^F$. In both
cases, the required result holds. 

The TBA analysis can be performed using the well known strategies. The only difficulty is the diagonalization of the monodromy matrix, which involves solving an auxiliary problem based on the $a_2^{(2)}$ vertex model. String solutions for this model were not known before, but they can easily be obtained using our recent results on the $a_2^{(1)}$ case. Setting $\gamma={\pi\over t-1}$, the $a_2^{(2)}$ Bethe equations have the form 
\begin{equation}
\prod_\alpha {\sinh{1\over 2}(y_i-u_\alpha-i\gamma)\over  \sinh{1\over 2}(y_i-u_\alpha+i\gamma)} =
\prod_j {\sinh{1\over 2}(y_i-y_j-2i\gamma)\over  \sinh{1\over 2}(y_i-y_j+2i\gamma)}{\sinh{1\over 2}(y_i-y_j+i\gamma)\over  \sinh{1\over 2}(y_i-y_j-i\gamma)}
\end{equation}
where the $y_i$ are Bethe roots, and the $u_\alpha$ are spectral parameter heterogeneities (corresponding to the rapidities of particles already present in the system). The solutions of these equations in the thermodynamic limit are as follows. The $y$'s can be $1,2,\ldots,t-1$ strings, or antistrings. In addition, it is possible to have a $t$ string centered on an antistring, or to have a complex of the form $y=y_r\pm {i\gamma\over 2}+i\pi$.

After the usual manipulations, one ends up with  equations for the pseudoenergies, that can be represented 
 using a  TBA diagram. The `left part' of the diagram corresponds to the following equations
\begin{equation}
{\epsilon_j(\theta)\over T}=\phi_{3}(\theta-\theta')\star \ln\left(1+e^{\epsilon_j(\theta')/T}\right)-\sum_{l=0}^{t-3}  \left(\delta_{j,l+1}+\delta_{j,l-1}\right)\phi_{3}(\theta-\theta')\star\ln\left(1+e^{-\epsilon_l(\theta')/T}\right)
\end{equation}
where we denote $\phi_P(\theta)={P\over 2\cosh (P\theta/2)}$, $f\star g(\theta)=\int_{-\infty}^\infty {d\theta'\over 2\pi} f(\theta-\theta')g(\theta')$. We use in the following the Fourier transform
\begin{equation}
\hat{f}(\omega)=\int {d\theta\over 2\pi} e^{iP\omega\theta/\pi} f(\theta)
\end{equation}
so 
$\widehat{(f \star g)}=2\pi \hat{f}\hat{g}$, 
and $\hat{\phi}_P={1\over 2\cosh\omega}$. 
We introduce the other kernel $\psi$ defined 
by $\hat{\psi}={\cosh \omega/2\over \cosh\omega}$.

In addition, there is a set of equations providing a closure on the right part. 
\begin{eqnarray}
{\epsilon_{t-3}(\theta)\over T}&=&\phi_{3}(\theta-\theta')\star \ln\left(1+e^{\epsilon_{t-3}(\theta')/T}\right)-  \phi_{3}(\theta-\theta')\star\ln\left(1+e^{-\epsilon_{t-4}(\theta')/T}\right)-\nonumber\\
&-&\sum_{i=1}^3  \phi_{3}(\theta-\theta')\star\ln\left(1+e^{-\epsilon_{a_i}(\theta')/T}\right)-\psi(\theta-\theta')
\star\ln\left(1+e^{-\epsilon_{b}(\theta')/T}\right)
\end{eqnarray}
Together with 
\begin{eqnarray}
{\epsilon_{a_i}(\theta)\over T}&=&-\phi_{3}(\theta-\theta')\star \ln\left(1+e^{-\epsilon_{t-3}(\theta')/T}\right)+
\phi_{3}(\theta-\theta')\star \ln\left(1+e^{\epsilon_{a_i}(\theta')/T}\right) \nonumber\\
&+&\sum_{j\neq i}  \phi_{3}(\theta-\theta')\star\ln\left(1+e^{-\epsilon_{a_j}(\theta')/T}\right)+\psi(\theta-\theta')
\star\ln\left(1+e^{-\epsilon_{b}(\theta')/T}\right)
\end{eqnarray}
and
\begin{eqnarray}
{\epsilon_{b}(\theta)\over T}&=&-\psi(\theta-\theta')\star \ln\left(1+e^{-\epsilon_{t-3}(\theta')/T}\right)+
2\phi_{3}(\theta-\theta')\star \ln\left(1+e^{\epsilon_{b}(\theta')/T}\right) \nonumber\\
&+&\sum_{i=1,2}  \psi(\theta-\theta')\star\ln\left(1+e^{-\epsilon_{a_i}(\theta')/T}\right)+\psi(\theta-\theta')
\star\ln\left(1+e^{\epsilon_{a_3}(\theta')/T}\right)
\end{eqnarray}
Finally, the asymptotic conditions
$\epsilon_0(\theta\rightarrow\infty)\rightarrow m\cosh\theta$ must be
imposed. This system can be conveniently encoded in the diagram of
figure 5. 
The free energy per unit length reads as usual
\begin{equation}
F=-T \int_{-\infty}^\infty {d\theta\over 2\pi}
m\cosh\theta\ln\left(1+e^{-\epsilon_0/T}\right)
\end{equation}

\begin{figure}
\centerline{\epsfxsize=3.0in\epsffile{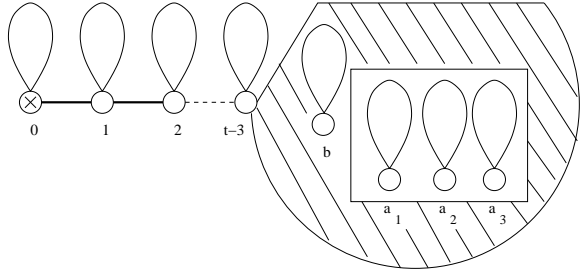}}
\bigskip\bigskip
\caption{Incidence diagram for the TBA of the (anisotropic)
  $a_2^{(2)}$ theory. Nodes are associated with the pseudoenergies
  $\epsilon$, and the cross indicates the presence of a massive
  asymptotic behavior for $\epsilon_0$.}
\end{figure}

We will  consider the more general case of  twisted boundary conditions, by adding a
phase factor in the trace  $Z= Tr \left[e^{-\beta H}e^{i \alpha q/(t-1)}\right]$, $q$ the topological charge. The
kinks have  therefore  a fugacity $ (e^{\pm i\alpha/t-1},1)$.  
We concentrate on the central charge, which is expressed in terms of 
the  quantities $x=e^{-\epsilon/T}$ in the limits of large and 
small temperature. At large temperature (UV), 
the $x_j$  go to constants, which solve the following system (here we
set $\lambda=e^{i\alpha}$, which appears in the equations due to a
renormalization of the spin \cite{FendleySaleur}):
\begin{eqnarray}
x_0=&(1+x_1)^{1/2} \left(1+{1\over x_0}\right)^{-1/2}\nonumber\\
\ldots &\nonumber\\
x_n=&(1+x_{n-1})^{1/2}(1+x_{n+1})^{1/2}\left(1+{1\over x_n}\right)^{-1/2}\nonumber\\
\ldots &\nonumber\\
x_{t-3}=&(1+ x_{t-4})^{1/2} (1+x_a)^{1/2}(1+\lambda x_a)^{1/2}(1+\lambda^{-1}x_a)^{1/2} \left(1+{1\over x_{t-3}}\right)^{-1/2} (1+x_b)\nonumber\\
x_a=&(1+x_{t-3})^{1/2} \left(1+{1\over x_a}\right)^{-1/2} (1+\lambda
x_a)^{-1/2}
(1+\lambda^{-1} x_a)^{-1/2} (1+x_b)^{-1}\nonumber\\
x_b=&(1+ x_{t-3})\left(1+{1\over x_a}\right)^{-1} (1+\lambda
x_a)^{-1}(1+\lambda^{-1} x_a)^{-1} 
(1+x_b)^{-1}
\end{eqnarray}
and recall that there are three (like the dimension of the fundamental representation) nodes with a common value of $x_a$. The solution  of this system is
\begin{eqnarray}
 x_j=&{ \sin {(j+1)\alpha\over 2t}\sin {(j+4)\alpha\over 2t}\over
   \sin{\alpha\over 2t} \sin{\alpha\over t}}&,\ \ j=0,\ldots,t-3\nonumber\\
x_a=&{\sin {(t-1)\alpha\over 2t}\over \sin {(t+1)\alpha\over 2t}}&\nonumber\\
x_b=&{\sin^2 {(t-1)\alpha\over 2t}\over \sin{\alpha\over t}\sin\alpha}&\
\end{eqnarray}
What we will in general call the UV contribution to the central charge is 
$c_1={6\over \pi^2}\sum L\left({x_j\over 1+x_j}\right)$. At small temperature (IR), the $x_j$ go similarly to constants solving the same system but with one less node on the left, because $x_0\rightarrow 0$. We first consider the case $\alpha=0$, i.e.  periodic boundary conditions for the bosons, antiperiodic boundary conditions for the fermions. In that case,  
the UV sum of dilogarithms gives  a contribution $(t-1)$, while one gets a  similar contribution from the IR sum
  after  $t\rightarrow t-1$, $c_2=t-2$. The central charge is thus $c=c_1-c_2=1$, as expected.

[Here we include two specialized remarks: 

A point of some interest is $\gamma={\pi\over 2}$, corresponding to $h_1={1\over 4}$. In that case, the $a_2^{(2)}$  Bethe equations 
do coincide (after a shift $y\rightarrow y+{i\pi\over 2}$) with  the $a_1^{(1)}$ Bethe equations that appear in solving the 
  sine-Gordon model 
with ${\beta_{SG}^2\over 8\pi}={1\over 4}$. This point is in the attractive regime, with one soliton and  one antisoliton of mass $m$, and one
 breather of the same mass. It is easy to check that in that case, the $a_2^{(2)}$ TBA is in fact identical with the well known SG TBA indeed.
 The equivalence between the two theories is not so obvious when one looks at the actions.

Also, as $a_2^{(2)}$ is related to $a_2^{(1)}$, so does the $a_2^{(2)}$ theory bear some resemblance to the $a_2^{(1)}$ Toda theory
 with the following action
\begin{equation}
S={1\over 8\pi} \int dx dy
\left[\sum_{i=1}^2(\partial_x\phi_i)^2+(\partial_y\phi_i)^2-
\Lambda'\left(e^{i{\beta\over \sqrt{2}}(\phi_1+\sqrt{3}\phi_2)}
+
e^{i{\beta\over \sqrt{2}}(\phi_1-\sqrt{3}\phi_2)}+
e^{-i\sqrt{2}\beta\phi_1}\right)\right] \label{suthree}
\end{equation}
 Here the perturbation has a single dimension
$h=\beta^2$, and the dimension of the coupling  is
$[\Lambda]=L^{2\beta^2-2}$.
Parameterizing $\beta$ in (\ref{suthree}) by   $\beta^2={t-1\over t}$, 
  it turns out that the free energy of the $a_2^{(2)}$ theory is exactly half the free energy of the $a_2^{(1)}$ theory, once the fundamental masses have been matched. This 
 fact does not appear obvious in the least when one compares 
 perturbative expansions!]

Twistings and truncations of the $a_2^{(2)}$ model are of the highest interest 
and have been widely discussed in the literature 
\cite{Smirnov, Warnaar, Takacs}.  
Twisting (that is, 
putting a charge at infinity) in such a way that    $e^{i\beta\phi}$ becomes a 
screening operator of weight $\Delta=1$, gives the central charge $c=1-{3(t-2)^2\over (t-1)t}$.
 RSOS restriction is then possible for $t$ even, giving rise to the minimal model 
$M_{t-1,t/2}$. The perturbation in
the minimal model has then weight $\Delta_{21}=1-{3\over 2t}$ (its coupling is real, and the sign does not matter because 
it has only even non vanishing correlators). Meanwhile, the lowest weight 
$\Delta_{12}={4-t\over 8(t-1)}$
  becomes negative for $t\geq 4$, after which the effective central
  charge reads $c_{\hbox{eff}}=c-24h_{12}=1-{12\over t(t-1)}$. 
One can also twist in such a way that $e^{-i\beta\phi/2}$ becomes a screening operator,
giving the central charge $c=1-{3(t+2)^2\over (t-1)t}$. RSOS restriction is then possible for $t$ odd,
giving rise to  $M_{t,(t-1)/2}$ perturbed by the operator of weight 
 $\Delta_{15}= 1-{3\over t}$. 
 \footnote{Notice that the combination $2-x={3\over t},\hbox{resp. }{6\over t}$ for $t$ even (resp. odd). In fact, the perturbative series for the free energy always has the same structure, and does not exhibit parity effects as $t$ is changed. But the physical interpretation does,
and rightly so, since for $\phi_{21}$, only even correlation functions do not vanish, while for  $\phi_{15}$, all correlation functions 
are a priori non vanishing.} 

The twisting can also be studied with  the TBA using now $\alpha\neq 0$. 
The UV sum of dilogarithms gives then  a contribution
$(t-1)-3{t-1\over t}
{\alpha^2\over\pi^2}$, while one gets a  similar contribution from the IR sum
  after  $t\rightarrow t-1$. The central charge is thus $c=1-{3\over
  t(t-1)}{\alpha^2\over\pi^2}$.

In general, twisting terms affecting nodes `far to the right' of the
TBA diagram do not affect  the central charge in the isotropic
limit. Indeed, if $\alpha$ were to remain finite here as $t\rightarrow
\infty$, the central charge of the twisted theory would be still
$c=1$. We shall however be interested in giving 
antiperiodic boundary conditions to the kinks of charge $q=\pm 1$,
which 
translates into  a phase   that blows up like $\alpha\approx t\pi$ 
 as $t\rightarrow
\infty$. As a result, the central charge of interest is $c\rightarrow
-2$, in agreement with the sigma model interpretation to be discussed
next.

\bigskip

Finally, we notice that choosing $\alpha=2\pi$ leads to $x_{t-4}=0$, 
and a truncation of the diagram to the one represented in figure 6.
This is the same as  folding the TBA for the $a_2^{(1)}$ RSOS model with central charge
 $c=2-{24\over (t-1)t}$. The first model in the series has $c_{\hbox{eff}}={2\over 5}$, the next one $c_{\hbox{eff}}={3\over 5}$ (the latter 
 TBA has some fascinating properties,
due to the fact that ${2\over 5}+{3\over 5}=1$). This was first
observed in \cite{Ravanini}. We will comment about the relation of
these models to $OSP(1/2)$ in the conclusion. 
For TBAs related with $a_2^{(2)}$ in other regimes, see \cite{RV, Mezincescu}.

\begin{figure}
\centerline{\epsfxsize=3.0in\epsffile{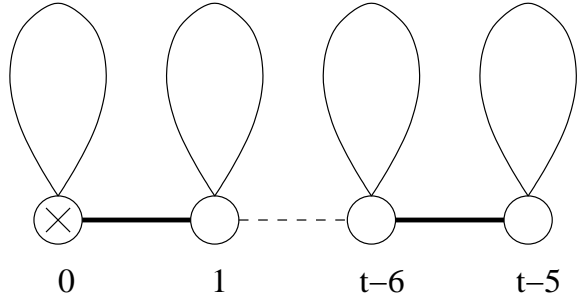}}
\bigskip\bigskip
\caption{Incidence diagram of the truncated $a_2^{(2)}$ TBA. This TBA
  also describes perturbations of the $OSP(1/2)_{(4-t)}/SU(2)_{(t-4)/2}$
  models for $t$ even, see later.}
\end{figure}

\subsection{The $OSP(1/2)$ limit, and the relation with the sigma model}

As explained previously, the $OSP(1/2)$ scattering theory can be studied by taking the $t\rightarrow\infty$ limit of the $a_2^{(2)}$ model.
The identification could in fact be seen directly 
 by identifying  Bethe equations. This seems a bit strange at first, because the $a_2^{(2)}$ equations do not have a structure that is reminiscent of the  $osp(1/2)$ Dynkin diagram. One has to remember however  
that the $osp(1/2)$ Bethe ansatz equations   are  peculiar, and their structure is not related with 
the Cartan matrix in the usual way. They read in fact  \cite{MartinsRamos}
\begin{equation}
\prod{\lambda_i-\mu_\alpha-i\over \lambda_i-\mu_\alpha+i}=\prod {\lambda_i-\lambda_j-2i\over\lambda_i-\lambda'+2i}\prod{\lambda_i-\lambda_j+i\over
\lambda_i-\lambda_j-i}
\end{equation}
and match the $a_2^{(2)}$ equations in the $t\rightarrow\infty$ limit, with   $y=\gamma\lambda$, $u=\gamma\mu$, $\gamma\rightarrow 0$.

 The Toda theory (\ref{Toda})
can then be rewritten in terms of a Dirac fermion as
\begin{equation}
S=\int {d^2x\over 2\pi}\left[\psi^\dagger_R\partial\psi_R+\psi^\dagger_L\bar{\partial}\psi_L
+\Lambda\left(\psi^\dagger_R\psi_L+\psi_R\partial\psi_R\psi^\dagger_L\bar{\partial}\psi^\dagger_L\right)\right]\label{untwistedpert}
\end{equation}
the perturbation is the sum of a term of dimension $h_1={1\over 2}$, 
and a term of dimension $h_2=2$ (the relative normalization 
between the two fermionic terms is irrelevant, since it can be 
adjusted by $\psi_R\rightarrow \lambda\psi_R$, 
$\psi_R^\dagger \rightarrow \lambda^{-1}\psi_R^\dagger$, 
or similarly for left fermions). It is likely 
that this model  could be directly  diagonalized using  
the coordinate Bethe ansatz,
like  the ordinary massive Thirring model, but we have not carried
out such a calculation. Conserved quantities can be found in terms of
the fermions; the first ones are 
$\psi^\dagger_R$, $\psi^\dagger_R\partial(\partial\psi^\dagger_R\psi_R)$...
 
The twisted theory meanwhile has  $c=-2$, $c_{\hbox{eff}}=1$, and the perturbations both acquire dimensions $(1,1)$. 
This can be identified with  a symplectic fermion theory with action \cite{Kausch}
 \begin{equation}
 S=\int {d^2x\over 2\pi}\left[\partial_\mu\eta_1\partial_\mu\eta_2+\Lambda'\partial_\mu\eta_1\partial_\mu\eta_2+\Lambda''\eta_1\eta_2
 \partial_\mu\eta_1\partial_\mu\eta_2\right]\label{twistedpert}
 \end{equation}
Here $\eta_1$ and $\eta_2$ are two fermionic fields with propagator, in the free theory, $<\eta_1(z,\bar{z})\eta_2(0)>=-\ln z\bar{z}$. 
Notice how non unitarity is manifest in (\ref{untwistedpert}) as well as (\ref{twistedpert}).

From the point of view of the twisted theory, the perturbation
involves two fields of weights $(1,1)$ 
which should be identified 
with $\phi_{21}$ and $\phi_{15}$ respectively, using fusion relations. That both fields appear is not unexpected, since the $c=-2$ point is a limit, and should have the characteristics of both $t$ even and $t$ odd. 

The identification of $\partial_\mu\eta_1\partial_\mu\eta_2$ with
$\phi_{21}$ can actually be completed accurately, by comparing the
four point functions as calculated in the fermion theory and the
minimal model using the Dotsenko Fateev general results
\cite{Saleurpolymers}. An interesting sign subtlety appears in that
case. Indeed,
$\partial_\mu\eta_1\partial_\mu\eta_2=\bar{\partial}\eta_1\partial\eta_2+\partial\eta_1\bar{\partial}\eta_2$,
and if we call this operator $O$, $<O(1)O(2)>=-{1\over
  z_{12}^2\bar{z}_{12}^2}$ because of anticommutation
relations. Hence, $\partial_\mu\eta_1\partial_\mu\eta_2$ should
actually be identified with $i\phi_{21}$. In fact, when one compares
the amplitude of the perturbation in the $a_2^{(2)}$ Toda theory and
the twisted version \cite{Fateev}, one finds that, with  the  usual
normalizations, $\Lambda$ positive gives rise to the amplitude of
$\phi_{21}$ being purely imaginary, that is the coefficient $\Lambda'$
in (\ref{twistedpert}) real. The sign of $\Lambda'$ is irrelevant, as
only terms even in $\Lambda'$ will appear in the perturbative
expansions of physical quantities. 

It would be very interesting to complete the identification of
$\phi_{15}$ with
$\eta_1\eta_2\partial_\mu\eta_1\partial_\mu\eta_2$, but we have not
finished this calculation. Note however that there is little doubt
this identification is correct, as there is no other object with the
right dimension and statistics in the symplectic fermion theory. 
Defining $O=\eta_1\eta_2\partial_\mu\eta_1\partial_\mu\eta_2$, one
finds $<O(1)O(2)>= {1+(\ln |z_{12}|^2)^2\over |z_{12}|^2}$. The massive
perturbation with $\phi_{15}$ is obtained with a coefficient that is
real and positive near $\beta^2=2$ \cite{Fateev}. Therefore, we expect
$\Lambda''$ in (\ref{twistedpert}) to be positive. Note that the
apparition of logarithms in the two point function of the perturbing
operator
makes  the field theory (\ref{twistedpert}) a bit problematic. Issues
of renormalizability arise in particular, and it is probably better to
think of (\ref{twistedpert}) as a sector of (\ref{untwistedpert})
rather than the defining theory. This is reflected in the structure of the TBA: although $c=-2$ can formally 
be obtained as the UV value of the central charge in the untwisted model, this value appears only after proper analytic continuation of the 
dilogarithms. Indeed, 
the fugacity  given to end nodes of the TBA diagram 
is  $e^{i\alpha}$, and as as $t\rightarrow\infty$,
$\alpha\approx t\pi$,  it  winds an
infinite number of times around the origin : in
practice, following the free energy  would presumably require following
analytic continuations on an infinity of different branches, a
difficult task at best. 

The fermions can  always be rescaled to bring the action into the form
 \begin{equation}
 S=\int {d^2 x\over 2\pi}\left[\partial_\mu\eta_1\partial_\mu\eta_2+ \Lambda\eta_1\eta_2
 \partial_\mu\eta_1\partial_\mu\eta_2\right]
 \end{equation}
where again the coupling $\Lambda$ is positive. We will now see how this related to the super sigma model. 

In general, the  coset space $OSP(m/2n)/OSP(m-1/2n)$ has dimensions $(m-1,2n)$ and can be interpreted as 
the supersphere $S^{m-1,2n}$ \cite{Z}.  The case of interest here is $m=n=1$, and  corresponds to the $S^{0,2}$ supersphere, parameterized by the coordinates
\begin{eqnarray}
x_1&=&1-{1\over 2}\eta_1\eta_2\nonumber\\
\xi_1&=&\eta_1\nonumber\\
\xi_2&=&\eta_2
\end{eqnarray}
such that $x_1^2+\xi_1\xi_2=1$. The action of the sigma model will generally be of the form 
$$
{1\over g}\left(\sum_{i=1}^m(\partial_\mu x_i)^2+\sum_{j=1}^n\partial_\mu \xi_{2j-1}\partial_\mu \xi_{2j}j\right)$$ 
(our convention is that 
the Boltzmann weight is $e^{-S}$). The beta function will be to first order $\beta\propto -(m-2n-2)g^2$, so for the region $m-2n<2$ in which we are interested, the model will be free in the UV and massive in the IR for a {\sl negative} coupling constant,
$g=-|g|$. In the $S^{0,2}$ case, this action therefore reads 
\begin{equation}
S=-{1\over |g|}\int d^2 x \left[\partial_\mu\eta_1\partial_\mu\eta_2-{1\over 2}\eta_1\eta_2 \partial_\mu\eta_1\partial_\mu\eta_2\right]
\end{equation}
Note that a rescaling combined  with a relabeling  can always bring this action into the form 
\begin{equation}
S=\int d^2 x \left[\partial_\mu\eta_1\partial_\mu\eta_2+{|g|\over 4\pi}\eta_1\eta_2 \partial_\mu\eta_1\partial_\mu\eta_2\right]
\end{equation}
matching the $t\rightarrow\infty$ limit of the $a_2^{(2)}$ theory,
with $\Lambda\propto |g|$.

\section{Supersphere sigma models and integrable superspin chains}

The relation we uncovered between $a_2^{(2)}$ and $OSP(1/2)$ extends immediately 
to the case of $a_{2n}^{(2)}$ and $OSP(1/2n)$: one can establish, for general values of $n$, the relation between the quantum affine algebras,
the Bethe ansatz equations, the scattering matrices etc. We thus propose that  the S matrix with $OSP(1/2n)$ symmetry,
represented in (\ref{maini}),(\ref{mainii}),(\ref{mainiii}) 
with $N=1-2n$, and the prefactor $\sigma_2^+$, 
provides an analytic continuation of the $O(N)/O(N-1)$ ``sphere'' sigma model
to this value of $N$. 

Of course, the analytic continuation of the sigma model should be interpreted as
the  coset $OSP(1/2n)/OSP(0/2n)$. The effective central charge of the UV limit 
is $c_{\hbox{eff}}=n$, while its true central charge will be $c=-2n$. For the ordinary sigma models, the UV central charge is $N-1$, 
so the UV value in the analytic continuation just matches. 

The  $a_{2n}^{(2)}$ Toda theory  has an interaction term of the form
$$
e^{{i\over \sqrt{2}}\beta(\phi_1-\phi_2)}+e^{{i\over \sqrt{2}}
\beta(\phi_2-\phi_3)}+\ldots+e^{{i\over
  \sqrt{2}}\beta(\phi_{n-1}-\phi_n)}+
e^{i\sqrt{2}\beta\phi_n}
$$
The dimension of vertex operators $\exp(i\sum\delta_j\phi_j)$ is
$h=\sum {1\over 2}\delta_j^2-\delta_0\sum\delta_j$, where $\delta_0$ measures the twist,
and the central charge is $c=n(1-12\delta_0^2)$. The operators in the
interaction term all have $h=1$ when $\delta_0={1\over 2}$, and
$\beta=\sqrt{2}$. In that case, $c=-2n$, while all the operators
$e^{\pm {i\over \sqrt{2}}
(\phi_j-\phi_k)}$ and $e^{2i\phi_j}$  have dimensions $(1,1)$.

The manifold relevant for $OSP(1,2n)$ on the other hand is $S^{0,2n}$, i.e. a purely ``fermionic sphere''. For instance, $S^{0,4}$ can be parameterized by
\begin{eqnarray}
x_1&=&1-{1\over 2}\left(\eta_1\eta_2+\eta_3\eta_4\right)-{1\over 4}\eta_1\eta_2\eta_3\eta_4\nonumber\\
\xi_1&=&\eta_1\nonumber\\
\xi_2&=&\eta_2\nonumber\\
\xi_3&=&\eta_3\nonumber\\
\xi_4&=&\eta_4
\end{eqnarray}
The action of the sigma model is not particularly illuminating; it involves four and six fermions couplings, and reduces to $2n$ symplectic fermions in the UV limit. Like in the $n=1$ case, it can be matched onto the appropriate limit of the $a_{2n }^{(2)}$ theory. 

On the other hand, it is also possible to extend the analysis of the $a_{2}^{(2)}$ TBA to arbitrary value of 
$n$, so we also know the TBA for this scattering theory, which is simply given by a $Z_2$ folding of the $a_{2n}^{(1)}$ TBA. 
The TBA is represented in figure 7.

\begin{figure}

\centerline{\epsfxsize=3.0in\epsffile{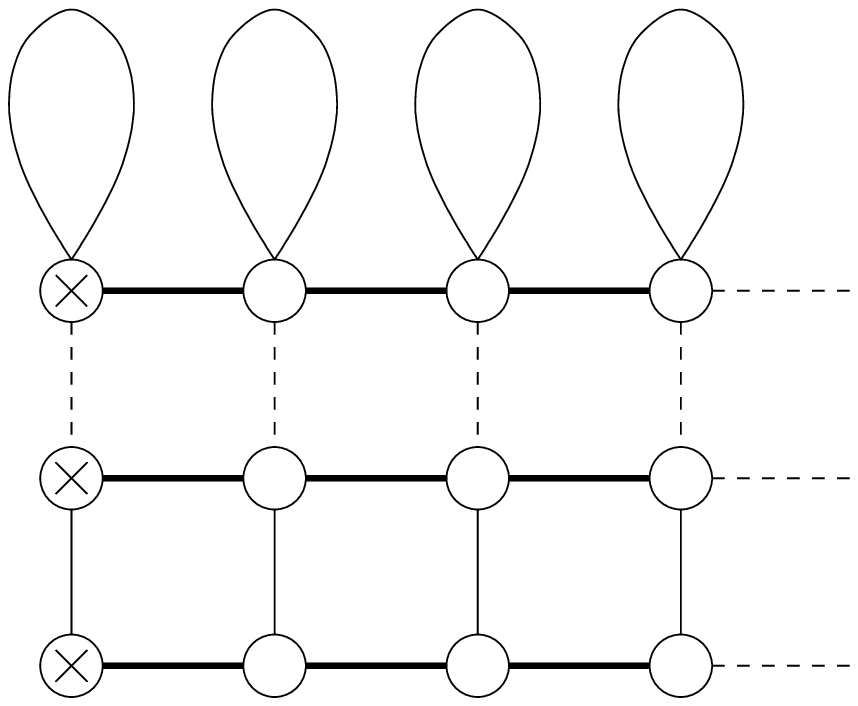}}
\bigskip\bigskip
\caption{Incidence diagram for  the TBA of  $OSP(1/2n)$ sigma
  models (the diagram has $n$ rows).}
\end{figure}

Notice that 
there are $n$ massive particles: while for $N$ integer positive the $O(N)/O(N-1)$ S matrix has no bound states, with simply $N$ fundamental particles (in the vector representation), poles do enter the physical strip for $N<2$. For the value $N=1-2n$ we are interested in, the masses of the particles 
are $m_i\propto \sin{i\pi\over 2n+1},\ i=1,\ldots,n$. The UV central charge is easily checked to be $c_{\hbox{eff}}=n$. We do not know how to obtain the central charge of 
the untwisted theory, as this would require a knowledge of the `closure' of the TBA diagram for twisted theories, an unsolved problem when $n>1$.

Our results have an immediate application to the study of quantum spin
chains. Indeed,
the Bethe equations which appear in the solution of the $OSP(1/2n)$
sigma models are similar to the ones appearing in the solution of the 
integrable $OSP(1/2n)$  chains studied in particular by Martins and
Nienhuis  \cite{MarNien}. More detailed calculations show that these
chains are critical, and that they coincide at large distance with the
weakly coupled supersphere sigma models, that is, a system of $2n$  free
symplectic fermions. This is disagreement with the conjecture in
\cite{MarNien,Martins} that this continuum limit should be a WZW model on the
supergroup: although the central charge agrees with both proposals,
detailed calculations of the thermodynamics or finite size spectra
show that the WZW proposal is not correct, and confirm the sigma model
proposal instead. A similar conclusion holds for $OSP(m/2n)$ when
$m-2n<2$. That the spin chain flows to the weakly coupled
sigma model is certainly related with the change of sign of the beta
function when $m-2n$ crosses the value 2, but we lack a detailed
understanding, similar to the ones proposed in \cite{Affleck,ReadSachdev},  of the mechanisms involved.

\section{The super Gross Neveu models}

\subsection{Generalities}

If we consider a scattering matrix defined again by 
(\ref{maini}),(\ref{mainii}),
but now with the prefactor $\sigma_2^-$ instead, it is natural to expect that it describes  $OSP(m/2n)$  Gross Neveu models,
the analytic continuation of the $O(N)$ GN models to $O(m-2n)$. 
Having a control on the diagonalization of $OSP(1/2n)$ scattering matrices  will allow us to study this scattering theory easily, and confirm the identification for these algebras. 
Notice that since  the $O(N)$  scattering matrix 
has no poles in the region $N<2$,  the roles of the GN and sigma models are  completely exchanged in the domain of values of $N$ we are considering. 

The  $OSP(m/2n)$ Gross Neveu models read
\begin{equation}
S=\int {d^2x\over 2\pi} \left[\sum_{i=1}^{m}\psi^i_L\partial\psi^i_L+\psi^i_R\bar{\partial}\psi^i_R+\sum_{j=1}^{n}
\beta_L^j\partial\gamma_L^j+\beta_R^j\bar{\partial}\gamma_R^j
+g
\left(\psi_L^i\psi_R^i+\beta_L^j\gamma_R^j-\gamma_L^j\beta^j_R\right)^2\right]
\end{equation}
This theory has central charge $c={m\over 2}-n$, effective central charge $c_{\hbox{eff}}={m\over 2}+2n$.  The beta function for this model is of the form
$\beta_g\propto (m-2n-2)g^2$, the same as the one for the $O(m-2n)$ GN model. For $m-2n>2$, it is thus positive,
so a positive  coupling  $g$ is marginally relevant - this is the usual massive GN model - while a negative one is marginally irrelevant. 
  If instead we consider the case $m-2n<2$, these results are switched: it is a negative coupling that is marginally relevant, and makes the theory  massive 
in the IR \footnote{In \cite{BL}, the four fermion coupling is defined through combinations $\bar{\psi}_-\psi_++\psi_-\bar{\psi}_+=2i (\psi_L^1\psi_R^1+\psi_L^2\psi_R^2)$, so what is called $g$ there is the opposite of our convention.}. The case $m=1$ should be described by the foregoing scattering theory. 

Note that the GN model is equivalent to the appropriate WZW model with a current current perturbation. Indeed, the system of $m$ Majorana fermions and $n$ symplectic bosons constitutes in fact a certain representation of the $OSP(m/2n)$ current algebra. The 
level depends on the choice of normalization; it would be called $k=1/2$ in \cite{Goddard}, $k=-1/2$ in \cite{ospwzw}, $k=1$ elsewhere. We adopt the latter convention here, and thus 
the level $k$ WZW model based on $OSP(m/2n)$ has central charge
\begin{equation}
c={(m-2n)(m-2n-1)k\over 2(m-2n-2+k)}
\end{equation}
Particular cases are $OSP(0/2n)$, which coincides with the $SP(2n)$ WZW model at level $-k/2$, and $OSP(m/0)$, which coincides with the 
$O(m)$ WZW model at level $k$. Super symmetric space theorems give rise to free fields representations at level $k=1$, where $c={m-2n\over 2}$, at level $k=m-2n-2$, where $c={(m-2n)(m-2n-1)\over 4}={\hbox{sdim } OSP(m/2n)\over 2}$.
 Notice that the representation at level $-2$ for $OSP(2/2)$ described  recently in \cite{andreas} is a particular case of the super symmetric space theorem discussed in Goddard et al. \cite{Goddard} (for $k=-1$ in their notations). 

The $OSP(m/2n)$ Gross Neveu models present additional non unitarity problems not encountered in the sigma models discussed above. To tackle these problems, we first discuss the simplest case of all. 

\subsection{The $OSP(0/2)$ case.}

We consider the case of the GN model for $N=-2$, corresponding formally to
$OSP(0/2)$, i.e. a $\beta\gamma$ system. 
The S matrix should act on a doublet of particles, and reads,
from the general formulas 
\begin{equation}
\check{S}= \tan\left({\pi\over 4}+{i\theta\over 2}\right)
{\Gamma\left({1\over 2}+{\theta\over 2i\pi}\right)\over
\Gamma\left({1\over 2}-{\theta\over 2i\pi}\right)}
{\Gamma\left(-{\theta\over 2i\pi}\right)\over
\Gamma\left({\theta\over 2i\pi}\right)}
\left(\begin{array}{cccc}
-1&0&0&0\\
0&{1\over \theta-i\pi}&-{\theta\over \theta-i\pi}&0\\
0&-{\theta\over \theta-i\pi}& {1\over \theta-i\pi}&0\\
0&0&0&-1
\end{array}\right)
\end{equation}
It turns out that $
\check{S}=i\tanh\left({\theta\over 2}-{i\pi\over 4}\right)
\check{S}_{SG}(\beta_{SG}^2=8\pi)$ where $S_{SG}$ is the soliton S matrix
of the sine-Gordon model. At coupling $\beta_{SG}^2=8\pi$, it coincides
with the S matrix of the $SU(2)$ invariant Thirring model, or the
level 1 WZW model with a current current perturbation. 

The scattering matrix is thus the same as the one for the $k=1$ $SU(2)$  WZW
model 
up to a CDD factor. This CDD factor does not introduce any additional
physical pole, but affects the TBA in an essential way. 

Note that the $N=-2$ Gross Neveu model can also be considered as a
current current perturbation of the $SU(2)$ WZW model at level
$k=-{1\over 2}$, or the $SU(1,1)$ model at level $k={1\over 2}$. 

\begin{figure}
\centerline{\epsfxsize=4.0in\epsffile{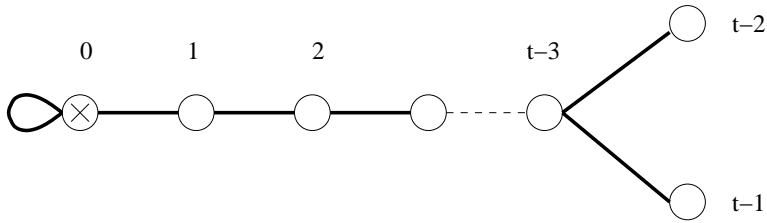}}
\bigskip\bigskip

\caption{Incidence diagram for the TBA describing the current-current perturbation of the (anisotropic)
  $\beta\gamma$
  system (or $OSP(0/2)$) model. }
\end{figure}

To study the TBA, it is useful as in the sigma model case to consider the anisotropic
deformation, with the sine-Gordon part now corresponding to
${\beta_{SG}^2\over 8\pi}={t-1\over t}$. The diagram is represented in figure
8.
The UV solutions have to obey the usual SG equations plus the
fact that 
\begin{equation}
x_0=(1+x_1)^{1/2}(1+x_0)^{1/2}
\end{equation}
The solution is obtained by setting $x_j=(j+\alpha)^2-1$,
$j=0,\ldots,t-3$, $x_{t-2}=x_{t-1}=t-3+\alpha$, and letting
$\alpha\rightarrow\infty$. The contribution to the central charge in
the UV is then $c_1=t$, the number of nodes. The solution in the IR
is obtained by discarding the first node, and then coincides with the
usual IR solution of the SG equations, with $\alpha=1$. The
contribution to the central charge is equal to $c_2=t-2$, the
number of nodes {\sl minus one}. The final central charge is thus
$c=2$, as expected for the effective central charge of the 
$\beta\gamma$ system.  

The same calculation with a fugacity $e^{i\pm {\alpha\over t-1}}$
 gives
 $c=2-{6\alpha^2\over (t-1)\pi^2}$. This is because in the UV, all the $x$'s are still infinite, giving rise this time to $c_1=t-{6\alpha^2\over\pi^2}$, while in the IR, the $x$'s are the same as the ones for the ordinary Sine-Gordon model, with $c_2=t-2-6{t-2\over t-1}{\alpha^2\over\pi^2}$. 
This result requires explanations; in particular, setting $\alpha=(t-1)\pi$ and letting $t\rightarrow\infty$ as in the sigma model case gives $c=-\infty$!

\subsection{The role of zero modes}

We want to  consider in more details the $\beta\gamma$ system with action
\begin{equation}
S=\int {d^2x\over 2\pi}
\left(\beta_L\partial\gamma_L+\beta_R\bar{\partial}\gamma_R\right)
\end{equation}
The propagators are $\gamma_R(z)\beta_R(w)=-\beta_R(z)\gamma_R(w)={1\over
  z-w}$. We can `bosonize' the ghosts by introducing a scalar field
  $\Phi=\phi_R+\phi_L$, such that $\phi_R(z)\phi_R(w)=-\ln(z-w)$. We
  also introduce fermionic ghosts
  $\eta_R(z)\xi_R(w)=\xi_R(z)\eta_R(w)={1\over z-w}$, and thus
\begin{eqnarray}
\gamma_R=e^{\phi_R}\eta_R,&\gamma_L=e^{-\phi_L}\eta_L\nonumber\\
\beta_R=e^{-\phi_R}\partial\xi_R,&\beta_L=e^{\phi_L}\bar{\partial}\xi_L
\end{eqnarray}
The corresponding action is then
\begin{equation}
S={1\over 8\pi}\int d^2x~(\partial_\mu\Phi)^2={1\over 2\pi}\int~d^2x~
\partial\Phi\bar{\partial}\Phi
\end{equation}
The $\beta\gamma$ 
Hamiltonian is 
\begin{equation}
H={1\over 4\pi}\int dx
\left(\beta_L\partial_x\gamma_L+\beta_R\partial_x\gamma_R\right)
\end{equation}
with commutators $[\beta_L(x),\gamma_L(y)]={i\over 4\pi}\delta(x-y)$,
$[\beta_R(x),\gamma_R(y)]=-{i\over 4\pi}\delta(x-y)$. 
The $U(1)$ current is given by $J_R=\gamma_R\beta_R=\partial\phi_R$,
$J_L=
\gamma_L\beta_L=-\partial\phi_L$. The topological charge is 
\begin{equation}
Q={1\over 2\pi}\int ~dx~(J_R-J_L)={1\over 2\pi}\int ~dx ~\partial_x\Phi
\end{equation}
The topological charge of $\gamma_L$ and $\gamma_R$ is 1, while the
charge of $\beta_L$ and $\beta_R$ is $-1$.  

A key feature of this system is the existence of zero modes. With
periodic boundary conditions, it is indeed easy to see that $[H,\int
dx\beta_{L,R}]=[H,\int dx\gamma_{L,R}]=0$. It follows from this that,
if we add to the Hamiltonian a term of the form $-hQ$, the system will
fill up with an infinity of zero mode particles of $\beta$ or $\gamma$
type depending on the sign of $h$, sending the ground state energy to
$-\infty$. The theory is thus unstable without a mass term. In the current algebra language, the 
infinite dimensional  space associated with the zero mode decomposes into lowest weight representations of $SU(1,1)_{1/2}$ of `angular momentum' $j=-{1\over 4}$ and $j=-{3\over 4}$. The conformal weight of these states is $\Delta=-{1\over 8}$, giving the effective central charge $c_{\hbox{eff}}=2$ for a $c=-1$ theory indeed. 

The mass term (which is actually  a current current perturbation) 
in the $OSP(0/2)$  GN model does stabilize the theory. 
To see how, let us add to the action a term 
\begin{equation}
\delta S=-{h\over 2\pi}\int d^2x (\gamma_R\beta_R-\gamma_L\beta_L)
+{g\over 8\pi}\int~ d^2x~ (\gamma_R\beta_L-\gamma_L\beta_R)^2\label{changeofs}
\end{equation}
The classical minima of $S+\delta S$  occur  for 
$\gamma_R=\gamma_L=c$ and $
\beta_R=-\beta_L=b$, and, turning to the Hamiltonian formalism, the minimum energy becomes then
\begin{equation}
{1\over L}E_{gs}=-{1\over 2\pi} {h^2\over g}\label{gsbega}
\end{equation}
We now recall the RG equation for the coupling constant $g$ in 
(\ref{changeofs}):
\begin{equation}
\dot{g}=-2g^2
\end{equation}
From this, the coupling constant at scale $1/m$ goes like $g={1\over
2\ln {cst/m}}$. The constant term is a UV cut-off, provided here by
the field $h$. It follows that
\begin{equation}
{1\over L}E_{gs}=-{1\over \pi}h^2\ln(h/m)
\end{equation}
at leading order. If $m\rightarrow 0$ ($g\rightarrow 0$), we recover
the result $E_{gs}\rightarrow -\infty$ anticipated before.

We will comment more on the behavior of the $OSP(0/2)$ and other GN
models later. For the moment, our goal is to explain the behavior of
the central charge in the anisotropic case obtained in the previous
section. So, we now consider  the case where an anisotropy is imposed on
the system by adding a coupling of the form $J_LJ_R$. More
explicitly,  consider 
\begin{equation}
\delta~A=-{h\over 2\pi}\int d^2x (\gamma_R\beta_R-\gamma_L\beta_L)
-{g\over 8\pi}\int~ d^2x ~2\gamma_R\beta_R\gamma_L\beta_L\label{blabla}
\end{equation}
It is easy to calculate the ground state energy at leading order as
$g\rightarrow 0$, which turns out to be
finite now, even though the theory is still massless:
$E_{gs}/L=-{h^2\over \pi g}$. Anisotropy has
stabilized the UV theory. 

We can now calculate this ground state energy using the
S matrix approach. To do so, we perturb the action by a term of the
form $\beta_R^2\gamma_L^2+\beta_L^2\gamma_R^2$. In the bosonized
version, this reads 
$$
e^{-2\Phi}\bar{\partial}\eta_L\eta_L\partial^2\xi_R\partial\xi_R
+e^{2\Phi}\partial\eta_R\eta_R\bar{\partial}^2\xi_L\bar{\partial}\xi_L
$$
The anisotropic term changes the kinetic term to  ${1\over
  8\pi}(1-{g\over 2}) (\partial_\mu\Phi)^2$. We can renormalize the
field so the kinetic term looks as before, and then the exponentials
in the perturbation become $e^{\pm 2\hat{\beta}\Phi}$ with
$\hat{\beta}^2={1\over 1+{g\over 2}}$. Non local
conserved currents are then obtained using
$$
\exp\left({-2\phi\over \hat{\beta}}\right) \partial^2\xi_R\partial\xi_R,~~~
\exp\left({2\phi\over \hat{\beta}}\right)\partial^2\eta_R\eta_R
$$
of dimension $\Delta_c=3-2/\hat{\beta}^2$. They lead to a quantum deformation of the $a_1^{(1)}$ algebra with
quantum parameter $q=e^{-i\pi\Delta_c}$. Setting $q=e^{i\pi {t-2\over
    t-1}}$, this corresponds to a thermodynamic Bethe ansatz diagram
with $t$ nodes (including the source one), i.\ e.\ the TBA studied in the
previous section and represented in figure 8. The point is, we now
have the correspondence between the coupling $g$ in the anisotropic
action and the parameter $t$ in the anisotropic TBA, with, 
at small $g$ or large $t$,
 $g\approx {1\over t}$. To use this TBA, we finally need to establish
 the correspondence between the magnetic field and the kinks
 fugacity: setting 
$e^{\pm h/T}=e^{i\pm \alpha/t-1}$,  where $T$ is the temperature, 
gives
$h={Ti\alpha\over t-1}$. Using this, the TBA result 
$c=2-{6\over\pi} {\alpha^2\over t-1}$ does match  the ground state
energy $E_{gs}=-{h^2\over\pi g}$ at leading order as $g\rightarrow
0$. We thus have explained the TBA results of the previous section in 
the light of the ground state instability of the $\beta\gamma$
system. 

For the
$\beta\gamma$ system itself, we of course obtain $c=-\infty$ for any non trivial
fugacity of the kinks. This can also be  understood as follows: the
partition function of the $\beta\gamma$ conformal field theory  with
periodic boundary conditions is infinite because of the presence of a
bosonic zero mode \cite{Guruswamyludwig}. On the other hand, in the
periodic sector, the TBA gives a finite result, with the central
charge $c={1\over 2}+2$. Therefore,  the TBA approach must describe
a renormalized partition function, divided by the infinite
contribution of the bosonic zero mode. In the antiperiodic sector
where there is no such zero mode, the result of this division is to
give zero, or, formally, a central charge equal to
$-\infty$. Nevertheless, the dependence of the ground state energy on
$\alpha$ can be predicted and checked using the TBA,
which provides another  non trivial check of the S matrix.

\subsection{The $OSP(1/2n)$ case}

We now get back to the $OSP(1/2n)$ case. The TBA turns out to have a
simple description in terms
of  $a_{2n}^{(2)}$  again. Consider therefore, not the $SU(2n+1)$ GN model, but a related scattering theory with only two multiplets of particles, corresponding 
respectively to the defining representation and its conjugate. Considering more generally the case of $SU(P)$ models, the relation 
between the $SU(P)$ GN scattering theory and this new theory is similar to the relation between the $O(P)$ GN model 
and the $O(P)/O(P-1)$ sigma model \cite{Paulletter}. We will  thus call this scattering theory `sigma model like', but we are not aware of any physical interpretation for it. The TBA equations can be written following the usual procedure. They read 
\begin{eqnarray}
{\epsilon_{aj}\over T}&=&\sum_{b=1}^{P-1}I_{ab}^{(P)} \phi_P\star\ln\left(1+e^{\epsilon_{bj}/T}\right)-
\sum_{l=1}^\infty I_{jl}^{(\infty)} \phi_P\star  \ln\left(1+e^{-\epsilon_{al}/T}\right),~~~j\geq 2\nonumber\\
{\epsilon_{a1}\over T}&=&\sum_{b=1}^{P-1}I_{ab}^{(P)} \phi_P\star\ln\left(1+e^{\epsilon_{bj}/T}\right)-
 \phi_P\star  \ln\left(1+e^{-\epsilon_{a2}/T}\right) \nonumber\\
 &-&(\delta_{a1}+\delta_{a,P-1})  \phi_P\star  \ln\left(1+e^{-\epsilon_{a0}/T}\right)
 \end{eqnarray}
for the pseudoparticles, and 
\begin{eqnarray}
{\epsilon_{10}\over T}&=&{m\cosh\theta\over T}+\sum_{a=1}^{P-1}\phi_{1a}\star\ln\left(1+e^{\epsilon_{a1}/T}\right)\nonumber\\
{\epsilon_{P-1,0}\over T}&=&{m\cosh\theta\over T}+\sum_{a=1}^{P-1}\phi_{P-1,a}\star\ln\left(1+e^{\epsilon_{a1}/T}\right)
\end{eqnarray}
In these equations again,   $\phi_P(\theta)={P\over 2\cosh{P\theta\over 2}}$, $\phi_{P-1,a}=\phi_{1,P-a}$, and $\hat{\phi}_{1a}(\omega)= {\sinh(P-a)\omega\over \sinh P\omega}$.

This TBA is in fact quite similar to the one of the ${\cal N}=2$ supersymmetric $SU(P)$ Toda theory \cite{FI} (the generalization of the supersymmetric sine-Gordon model for $SU(2)$): the difference affects only which nodes correspond to massive particles, and which ones to pseudo particles. As a result,
the central charge is easily determined, $c=2P-1$. Getting back to the
particular case $P=2n+1$, we can then fold this system to obtain (see
the appendix for the proof) the TBA for the $OSP(1/2n)$ Gross Neveu model,
whose  effective central charge  reads therefore  $c_{\hbox{eff}}={1\over 2} \left(2(2n+1)-1\right)=2n+{1\over 2}$. 

As an example, we can discuss in more details the case of the $OSP(1/2)$ GN model 
whose TBA is represented in the figure 9. We will even consider an
anisotropic generalization of this TBA, where the diagram is truncated
to the right in a way that is equivalent to what happened in the sigma
model case.

 There is now a single color index, and we relabel
 $\epsilon_{a,j}=\epsilon_j$.   Introducing $x_j=e^{-\epsilon_j/T}$,
 the equations in the UV are the same as for the $a_2^{(2)}$ Toda theory, except for the first one
 that reads now simply
\begin{equation}
 x_0=1+ x_1
\end{equation}
The closure equations in particular are the same as for the
$a_2^{(2)}$ anisotropic model. 

\begin{figure}
\centerline{\epsfxsize=4.0in\epsffile{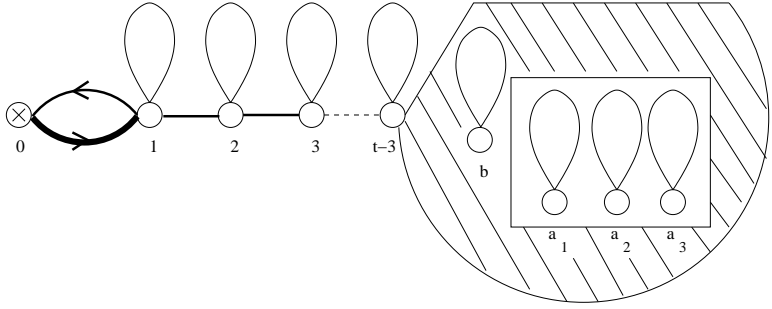}}
\bigskip\bigskip
\caption{Incidence diagram for the TBA describing the anisotropic version of the $OSP(1/2)$ GN model. }
\end{figure}

The solution in the case of a vanishing twist first is 
 $x_j={ (j+\alpha)(j+\alpha+3)\over 2}$, in the limit
 $\alpha\rightarrow\infty$, for $j=0,\ldots,t-3$. In addition one has
 $x_a={t-2+\alpha\over t+\alpha}$ and $x_b={(t-2+\alpha)^2\over
 4(t-1+\alpha)}$. As $\alpha\rightarrow\infty$, all the $x's$ go to
 infinity but $x_a$ which goes to one. As a result, the UV contribution to the central charge is $c_1=t-1+ 3\times {1\over 2}=t+{1\over 2}$.  
In the IR, the modification due to the $\sigma_2$ factor is not seen any longer, and the $x$'s obey the same equations as for the $a_2^{(2)}$ case, with a contribution $t-2$ to the central charge. It follows that $c={5\over 2}$, as expected. 

The twisted TBA follows from similar principles as in the sigma model
case. This time however, because of $\alpha=\infty$, the UV values
$x_j$  are unaffected by the twisting. The contribution to the central
charge is $c=t-1+{1\over 2}+
\left(L_\lambda+L_{\lambda^{-1}}\right)(x_a=1)=t+{1\over
  2}-{3\alpha^2\over\pi^2}$. The IR values do depend on $\alpha$,  with
formulas identical to  the sigma model case, and a contribution
$c=t-2-3{t-2\over t-1}{\alpha^2\over\pi^2}$. The resulting  central charge is
$c={5\over 2}-3{\alpha^2\over (t-1)\pi^2}$. The dependence on $\alpha$ is similar to what we observed in the $OSP(0,2)$ case,
for similar reasons. The factor $3$ in this formula, as opposed to the factor $6$ in the $OSP(0,2)$ case, has its origin
in the different relations between  the physical  anisotropy and the parameter $t$ in the TBA: in the $OSP(1/2)$ case, $g\approx {1\over 2t}$.

\section{Finite field calculations}

To give further evidence for our S matrices, we now present some 
finite field calculations. The idea, which has been 
worked out in great details in other cases \cite{Hasenfratz}, is to
compare S matrix and perturbative calculations for the ground state
energy of the theory in the presence of an external field. The S
matrix calculations are considerably simpler than the TBA ones
because, for a proper choice of charge coupling to the external field
the ground state fills up with only one type of particles, with
diagonal scattering, and the Wiener-Hopf method can be used to solve
the integral equations analytically. 

The S matrix calculations are very close to the ones already performed for
the $O(N)$ sigma model and the $O(N)$ Gross Neveu models. In fact, in
the region $m-2n\geq 2$, the calculations are identical, since the S
matrix elements are obtained by continuation $N\equiv m-2n$, and, in
the domain $m-2n\geq 2$, the integral representations are obtained by
the same continuation as well. For these cases, one thus immediately
checks that the continuation of the S matrix to $N=m-2n$ matches the
beta functions of the sigma or Gross Neveu models, which are, too,
obtained by this continuation.

Things are more interesting in the case $m-2n<0$ in particular, to
which we turn now. We consider first the $OSP$ sigma model, with $S$ matrices determined 
by $\sigma^+$. If we couple the external field to a charge of the form
\begin{equation}
Q\propto\int (x_1\partial_t x_2-x_2\partial_t x_1) dx\label{firstk}
\end{equation}
the ground state fills up with bosonic particles of the form $|1>\!+i|2>$, with diagonal scattering 
$s=\sigma_2^++\sigma_3^+(\theta)$. If meanwhile we couple the external field to a  charge of the form
\begin{equation}
Q\propto\int (\xi_1\partial_t \xi_2+\xi_2\partial_t \xi_1) dx\label{seck}
\end{equation}
the ground state fills up with fermionic particles of the form $|1>\!+|2>$, with diagonal scattering $s=\sigma_2^+-\sigma_3^+$.

Let us consider this latter  case. Using formulas given in the first
section,
one finds
\begin{equation}
\sigma_2^+-\sigma_3^+= {\Gamma(1+x)\over \Gamma(1-x)}{\Gamma(1/2-x)\over \Gamma(1/2+x)}
{\Gamma(1/2+\Delta+x)\over \Gamma(1/2+\Delta-x)}{\Gamma(1+\Delta-x)\over \Gamma(1+\Delta+x)}
\end{equation}
where $x={i\theta\over 2\pi}$ and $\Delta={1\over N-2}$. This turns out to coincide with 
\begin{equation}
\sigma_2^-+\sigma_3^-= {\Gamma(1+x)\over \Gamma(1-x)}{\Gamma(1/2-x)\over \Gamma(1/2+x)}
{\Gamma(1/2-\Delta+x)\over \Gamma(1/2-\Delta-x)}{\Gamma(1-\Delta-x)\over \Gamma(1-\Delta+x)}
\end{equation}
after a continuation $\Delta\rightarrow -\Delta$. Similarly, 
\begin{equation}
\sigma_2^--\sigma_3^-= {\Gamma(1+x)\over \Gamma(1-x)}{\Gamma(1/2-x)\over \Gamma(1/2+x)}
{\Gamma(1/2-\Delta+x)\over \Gamma(1/2-\Delta-x)}{\Gamma(-\Delta-x)\over \Gamma(-\Delta+x)}
\end{equation}
does coincide with 
\begin{equation}
\sigma_2^++\sigma_3^+= {\Gamma(1+x)\over \Gamma(1-x)}{\Gamma(1/2-x)\over \Gamma(1/2+x)}
{\Gamma(1/2+\Delta+x)\over \Gamma(1/2+\Delta-x)}{\Gamma(\Delta-x)\over \Gamma(\Delta+x)}
\end{equation}
after the same continuation.

This means that, in a  TBA calculation,  the ground state energy of
the 
$OSP(m/2n)$ sigma model
coupled to a fermionic charge follows from known calculations about the $O(N)$ 
Gross Neveu model after formally setting $N=m-2n$ {\bf and} performing a continuation $N-2\rightarrow 2-N$. From 
expressions in \cite{Hasenfratz} we find therefore, for $m-2n<0$
\begin{eqnarray}
E(h)-E(0)=-{h^2\over 2\pi} \left[1+{1\over N-2} {1\over \ln(h/m)}
-
\left({1\over N-2}\right)^2 {\ln\ln (h/m)\over \ln^2(h/m)}\right.\nonumber\\
\left.+ {1\over N-2}{C_N\over \ln^2(h/m)}+
O\left({\ln\ln (h/m)\over \ln^3(h/m)}\right)\right]
\end{eqnarray}
with $C_N=\ln\Gamma\left(1+{1\over N-2}\right)-\left(1-{1\over N-2}\right)\ln 2+1$,
and $N=m-2n$. From this we deduce \cite{Hollowoodff} the ratio of the first two coefficients of the beta function
as ${\beta_2\over \beta_1^2}={1\over N-2}$.  

Similarly, from TBA calculations, the ground state energy of the $OSP(m/2n)$ Gross Neveu model
follows from known calculations about the $O(N)$ sphere sigma model 
after formally setting $N=m-2n$ {\bf and} performing a continuation $N-2\rightarrow 2-N$. From 
expressions in \cite{Hasenfratz} we find therefore for $m-2n< 2$

\begin{equation}
E(h)-E(0)=(N-2){h^2\over 4\pi} \left[\ln(h/m)-
{1\over N-2} \ln\ln (h/m)+D_N+
O\left({\ln\ln (h/m)\over \ln(h/m)}\right)\right]
\end{equation}
where $D_N=-{3\over N-2}\ln 2-\left({1\over 2}+{1\over N-2}\right)-\ln\Gamma\left(1-{1\over N-2}\right)$. 
From this we deduce the ratio ${\beta_2\over \beta_1^2}=-{1\over
  N-2}$. Observe that the leading term follows from the calculations
for the $\beta\gamma$ system described in the previous section; all
that 
has to be changed is the beta function for the coupling $g$ in (\ref{gsbega}), 
 resulting in $E_{gs}=-{N-2\over 4\pi}h^2\ln(h/m)$ indeed. Remarkably,
 it is the ground state of the Gross Neveu model that has a leading
 $h^2\ln(h/m)$ dependence, while the ground state of the sigma model
 has a leading pure  $h^2$ dependence: the roles of Gross Neveu and
 sigma model are therefore switched compared to the usual $O(N)$
 situation.

The calculation with a coupling to the first kind of charge
(\ref{firstk}) for $m-2n<2$ or the second type of charge (\ref{seck}) for 
$m-2n>2$ poses difficulties, as the kernel does not factorize in the usual way then, so the Wiener Hopf method does not seem 
applicable.

Now recall that for the usual $O(N)$ sphere sigma model, ${\beta_2\over \beta_1^2}={1\over N-2}$,
and for the usual $O(N)$ Gross Neveu model, ${\beta_2\over \beta_1^2}={-1\over N-2}$. The ratios we found 
are thus the analytic continuations to $N\rightarrow m-2n$, as
desired\footnote{The existence of different  sectors in the
  $OSP(m/2n)$ models does not spoil this conclusion, as the beta
  functions 
are only trivially affected by the twists.}.

\section{Conclusions and speculations}

To conclude, although more verifications ought to be carried out to complete our identifications, we believe we 
have determined the scattering matrices for the massive regimes of the 
$OSP(m/2n)$ GN and the $OSP(m/2n)/OSP(m-1/2n)$ sigma models in the
simple case $m=1$, based on algebraic considerations as well as
thermodynamic Bethe ansatz calculations.

\smallskip

It is tempting to expect that at least some of our  results generalize to  other cases $OSP(m/2n)$ for $m>1$ and $m-2n<2$.  In all these cases, we expect that the S matrix of the sphere sigma model will be obtained from the  conjecture at the beginning of this paper,  with $N=m-2n$, for $N<2$.
The S matrix of the GN model is probably more complicated. Recall that in the case $N\geq 2$, it is given by the general conjecture only for $N>4$. When $N\leq 2$, we think it is probably given by the conjecture only for $N<0$.

\smallskip

\smallskip

Observe now that for the usual $O(N)$ case, the factors $\sigma_2^+$ and $\sigma_2^-$
do not exhibit poles and are equal for $N=3,4$. For these values, the
(unique) S matrix based on the general conjecture (\ref{maini}) describes correctly the sigma model. As for the
Gross Neveu model, its description is more subtle: it turns out
that the vector particles are actually unstable, and that the spectrum
is made of kinks only. 

In the case $N<2$ of interest here, the factors $\sigma_2^+$ and $\sigma_2^-$ similarly do not exhibit
poles and are equal for $N=1,0$. These cases would correspond for
instance to $OSP(3/2)$ and $OSP(2/2)$ respectively. It is very likely
that there again, the S matrices describe the sigma model, and not the
Gross Neveu model, for which the proper particle content has still to
be identified.

The $OSP(2/2)$ case is particularly intriguing. The  $S^{1,2}$ sphere can be  parameterized by 
\begin{eqnarray}
x_1&=&\cos\phi(1-{1\over 2}\eta_1\eta_2)
\nonumber\\
x_2&=&\sin\phi(1-{1\over 2}\eta_1\eta_2)
\nonumber\\
\xi_1&=&\eta_1\nonumber\\
\xi_2&=&\eta_2
\end{eqnarray}
The action of the sigma model now reads
\begin{equation}
S=-{1\over |g|}\int d^2 x \left[(\partial_\mu\Phi)^2(1-\eta_1\eta_2)+\partial_\mu\eta_1\partial_\mu\eta_2-\eta_1\eta_2 \partial_\mu\eta_1\partial_\mu\eta_2\right]
\end{equation}
where $\Phi$ is compactified, $\Phi\equiv \Phi \hbox{ mod } 2\pi$. A rescaling and a relabeling brings it into the form
\begin{equation}
S=\int d^2 x \left[-(\partial_\mu\Phi)^2(1+|g|\eta_1\eta_2)+\partial_\mu\eta_1\partial_\mu\eta_2+|g|\eta_1\eta_2
\partial_\mu\eta_1\partial_\mu\eta_2\right]
\end{equation}
with now $\Phi\equiv \Phi \hbox{mod } {2\pi\over \sqrt{|g|}}$.
We see in particular that in the limit $g\rightarrow 0$, the action is
simply the one of a free uncompactified boson and a free fermion, of
total central charge $c=-1$, and that the boson has a {\sl negative}
coupling : this system therefore coincides with the standard
``bosonization'' of the $\beta \gamma$ system in the limit
$|g|\rightarrow 0$! 

Notice that as soon as $m>1$, the negative sign of
the bosons coupling for $OSP(m/2n)$ sigma models will have to be
handled carefully; presumably the $OSP(2/2)$,  $\beta\gamma$ example
will provide a good example of how to do this.

\smallskip
Finally, we discuss the boundary value $N=2$, which exhibits some exceptional features. Indeed, 
from the integrable point of view,  the solution of the Yang Baxter
equation (combined with crossing and unitarity)  based on the generic
 S matrix is not unique  for $N=2$ (in contrast with the  other values of $N$) but admits one continuous
parameter $\gamma$. This solution is close to the sine-Gordon solution, and is
related to it by 
\begin{eqnarray}
S=\sigma_2+\sigma_3\nonumber\\
S_T=\sigma_1+\sigma_2\nonumber\\
S_R=\sigma_1+\sigma_3
\end{eqnarray}
where $S,S_T,S_R$ are the usual sine Gordon amplitudes
\begin{eqnarray}
S_T&=&-i {\sinh (8\pi\theta/\gamma)\over \sin
  (8\pi^2/\gamma)}\nonumber\\
S&=&-i {\sinh (8\pi(i\pi-\theta)/\gamma)\over \sin
  (8\pi^2/\gamma)}S_R\nonumber\\
S_R&=&{1\over\pi} \sin (8\pi^2/\gamma) U(\theta)
\end{eqnarray}
and $U(\theta)$ is given e.g. in \cite{ZamoZamo}. 

In the case of $O(2)$, the existence of this parameter corresponds to
the fact that the $O(2)/O(1)$ sigma model, or the $O(2)$ GN model are
actually massless critical theories, the coupling $g$ being exactly
marginal. The S matrices then provide a massless description of these
theories. Since $\sigma_2^+=\sigma_2^-$, the $O(2)/O(1)$ sigma model
and the $O(2)$ GN model coincide; their identity follows from
bosonization of the massless Thirring model into the Gaussian model. 
The free parameter in the S matrices is related with the coupling
constant in either version of the model. 
(Note that the S matrices can also be used to describe some massive
perturbations. These, however,  give rise  to  different  type of models
than the ones  we are interested in, like the massive Thirring model. ) 

It seems very likely that similar things occur for $OSP(2n+2/2n)$
models as well. The identity of the sigma model and the GN model in
that case is not obvious, but one can at least check using our general
formulas that the central charge and the effective central charge do
match, $c_{\hbox{eff}}=3n+1$, $c=1$. 
There are on the other hand strong arguments showing that the
beta function is exactly zero \cite{Wegner}, so these models should have a line of
fixed points indeed \cite{Be}, in agreement with the S matrix prediction.

\bigskip
Besides completing the identifications we have sketched here, the most pressing questions that come to mind
are: what are the $S$ matrices of the Gross Neveu models for non generic values of $N$, 
what are the $S$ matrices for the multiflavour GN models, what are the $S$ matrices for the orthosymplectic Principal Chiral Models?
We hope to report some answers to these questions soon.  As a final related remark, 
recall \cite{embedding} that there is an embedding \footnote{The level $-2k$ in this formula stems from our conventions; it would be $k$ if it were defined 
with respect to the sub $SU(2)$.}
\begin{equation}
OSP(1/2)_{-2k}\approx SU(2)_k \times {OSP(1/2)_{-2k}\over SU(2)_k}
\end{equation}
and that the branching functions of the latter part define a Virasoro
minimal model, with 

\begin{eqnarray}
c_{osp}&=&{2k\over 2k+3}\nonumber\\
c_{su2}&=&{3k\over k+2}\nonumber\\
c_{virasoro}&=&1-6{(k+1)^2\over (k+2)(2k+3)}
\end{eqnarray}
For $k$ an integer, the situation is especially interesting. The
Virasoro models which appear there have $p=2k+3,q=k+2$; they are non
unitary, 
and their effective central charge is $c_{\hbox{eff}}=1-{6\over
  (k+2)(2k+3)}$. These models can thus be considered as $UOSP/SU$
coset models. Their perturbation by the
operator $\phi_{21}$ with dimension $h=1-{3\over 4(k+2)}$ coincides
with the RSOS models defined in section 3 as truncations of the
$a_2^{(2)}$ theories with $t=2k+4$.  We thus see that the supersphere sigma model
appears as the limit $k\rightarrow\infty$ of a series of coset models \cite{Paulletter},
just like the ordinary sphere sigma model say appears as the limit of
a series of parafermion theories, this time of $SU(2)/U(1)$
type. There are many other interesting aspects of $OSP$ coset models
in relation with the present paper which we also plan to discuss elsewhere.

\bigskip
\noindent{\bf Acknowledgments:} This work was supported by the DOE and
the NSF. HS thanks IPAM at UCLA where part of this work was done. 
B.W.-K. acknowledges support from the Deutsche Forschungsgemeinschaft 
(DFG) under the contract KA 1574/1-2.
We thank D. Bernard, M. Grisaru, P. Mathieu, S. Penati,  A. Tsvelik 
and especially N. Read and G. Takacs for
discussions.

\appendix

\section{Folding the $SU(P)$ TBAs}

The quantization equations for the $SU(P)$ GN model have been written for instance in \cite{Paul}:
\begin{eqnarray}
2\pi P_{a0}&=&m_a\cosh\beta+\sum_{b=1}^{P-1}
Y_{ab}^{(P)}
\star\rho_{b0}-\sum_{j=1}^\infty
\sigma_j^{(\infty)}\star\tilde{\rho}_{aj}, a=1,\ldots,P-1\nonumber\\
2\pi\rho_{aj}&=&\sigma_j^{(\infty)}\star \rho_{a0}-\sum_{b=1}^{P-1}\sum_{l=1}^\infty
A_{jl}^{(\infty)}\star K_{ab}^{(P)}\star\tilde{\rho}_{bl}\nonumber\\
\end{eqnarray}

Using  Fourier transform: $\hat{f}(\omega)=
 \int_{-\infty}^\infty {d\omega\over 2\pi} e^{Pi\omega\beta/\pi} f(\beta)$, one then has
\begin{equation}
\hat{Y}_{ab}^{(P)}=\delta_{ab} -e^{|\omega|} {\sinh((P-a)\omega)\sinh(b\omega)\over \sinh(P\omega)\sinh(\omega)}, a\geq b
\end{equation}
with $\hat{Y}_{ab}=\hat{Y}_{ba}$. The $Y^{(P)}$ kernels are logarithmic derivatives of the scattering matrix between top components in each of the fundamental representations. The group structure is encoded in the densities $\rho$ of (massless) pseudoparticles, which appear in the solution of the auxiliary Bethe system diagonalizing the monodromy matrix.

We now set  $P=2n+1$. The folded  equations then read
\begin{eqnarray}
2\pi P_{a0}&=&m_a\cosh\beta+\sum_{b=1}^n \left(Y_{ab}^{(2n+1)}+Y_{a,2n+1-b}^{(2n+1)}\right) \star\rho_{b0}-\sum_{j=1}^\infty \sigma_j^{(\infty)}\star\tilde{\rho}_{aj},a=1,\ldots,n\nonumber\\
2\pi\rho_{aj}&=&\sigma_j^{(\infty)}\star \rho_{a0}-\sum_{b=1}^{n}\sum_{l=1}^\infty A_{jl}^{(\infty)}\star\left(K_{ab}^{(2n+1)}+
K_{a,2n+1-b}^{(2n+1)}\right) \star\tilde{\rho}_{bl}
\end{eqnarray}
For instance,  $Y_{11}^{(2n+1)}+Y_{1,2n}^{(2n+1)}=-e^{(-n+1/2)|\omega|}{\sinh|\omega|\over \cosh (n+1/2)|\omega|}$. It is easy to check that the 
corresponding kernel  coincides with $\sigma_3^+-\sigma_2^+$ for $N=1-2n$, and 
with the corresponding $S$ matrix element in the $a_{2n}^{(2)}$ scattering theory \cite{Gandenberger}. The couplings between pseudoparticles 
can also be checked to arise from the structure of solutions of the $a_{2n}^{(2)}$ Bethe equations, generalizing the $a_2^{(2)}$ case.

The `sigma model like' equations for $SU(P)$ are based on a hypothetical scattering theory 
with physical   particles in the vector representation and its conjugate only. They read

\begin{eqnarray}
2\pi P_{10}&=&m\cosh\beta+ Z_{11}^{(P)}\star\rho_{10}+ Z_{1,P-1}^{(P)}\star\rho_{P-1,0}-\sum_{j=1}^\infty \sigma_j^{(\infty)}\star\tilde{\rho}_{1j}\nonumber\\
2\pi P_{P-1,0}&=&m\cosh\beta+ Z_{1,P-1}^{(P)}\star\rho_{10}+ Z_{P-1,P-1}^{(P)}\star\rho_{P-1,0}-\sum_{j=1}^\infty \sigma_j^{(\infty)}\star\tilde{\rho}_{P-1,j}\nonumber\\
2\pi\rho_{aj}&=&\sigma_j^{(\infty)}\left(\delta_{a1}+\delta_{a,P-1}\right)\star \rho_{a0}-\sum_{b=1}^{P-1}\sum_{l=1}^\infty A_{jl}^{(\infty)}K_{ab}^{(P)}\star \star\tilde{\rho}_{bl}
\end{eqnarray}
where $\hat{Z}_{11}^{(P)}=\hat{Z}_{P-1,P-1}^{(P)}= e^{-|\omega|}{\sinh ((P-1)\omega)\over\sinh(P\omega)}$ and 
$\hat{Z}_{1,P-1}^{(P)}=\hat{Z}_{P-1,1}^{(P)}= e^{-|\omega|}{\sinh (\omega)\over\sinh(P\omega)}$. The kernel $Z_{11}^{(P)}$ is the logarithmic 
derivative of what is called $F_{min}^{VV}$ in \cite{Paul}, ${1\over i}{d\over d\beta}\ln F_{min}^{VV}(\beta)=Z_{11}^{(P)}$. Setting again $P=2n+1$ and folding gives now 
\begin{eqnarray}
2\pi P_{10}&=&m\cosh\beta+ \left(Z_{11}^{(P)} +Z_{1,P-1}^{(P)}\right)\star\rho_{1,0}-\sum_{j=1}^\infty \sigma_j^{(\infty)}\star\tilde{\rho}_{1j}\nonumber\\
2\pi\rho_{aj}&=&\sigma_j^{(\infty)}\delta_{a1}\star \rho_{a0}-\sum_{b=1}^{n}\sum_{l=1}^\infty A_{jl}^{(\infty)}
\star\left(K_{ab}^{(2n+1)}+
K_{a,2n+1-b}^{(2n+1)}\right)\star\tilde{\rho}_{bl}
\end{eqnarray}
The kernel $\hat{Z}_{11}^{(2n+1)}+\hat{Z}_{1,2n}^{(2n+1)}={\sinh (2n \omega)+\sinh \omega\over \sinh (2n+1)\omega} e^{-|\omega|}$. It differs from the previous kernel $\hat{Y}_{11}^{(2n+1)}+\hat{Y}_{1,2n}^{(2n+1)}$
by $-{\cosh {2n-3\over 2}\omega\over \cosh {2n+1\over 2}\omega}$, which coincides with the Fourier transform of the 
 ratio $\sigma_2^+\over \sigma_2^-$ for $N=1-2n$. 

Some integral representations to finish (used in the domain $N\leq 0$)
\begin{equation}
\ln \sigma_2^-=\int_0^\infty \left(e^{i(2-N)\beta\omega/\pi}+ e^{-(2-N)\omega}e^{-(2-N)i\omega \beta/\pi}\right) {e^{-2\omega}-1\over e^{-(2-N)\omega}+1} {d\omega\over \omega}
\end{equation}
and 
\begin{equation}
\ln {\sigma_2^+\over \sigma_2^-}=\int_{-\infty}^\infty e^{i(2-N)\beta \omega/\pi} {\cosh {N+2\over 2} \omega\over \cosh {N-2\over 2}\omega} {d\omega\over \omega}
\end{equation}
where $\sigma_2^+={\sinh \theta-i\sin{2\pi\over N-2}\over \sinh\theta+i\sin{2\pi\over N-2}}\sigma_2^-$.

\end{document}